\documentclass[sigconf]{acmart}
\usepackage{amsmath} 
\usepackage[font=small,skip=2pt]{caption}
\usepackage{colortbl,acmart-taps}  
\usepackage{dblfloatfix} 
\usepackage{graphics} 
\usepackage{multirow} 
\usepackage{soul} 
\usepackage[font=small,skip=1pt]{subcaption} 
\usepackage{titlesec} 
\usepackage{xspace} 

\newcolumntype{?}{!{\vrule width 1pt}}
\newcolumntype{C}[1]{>{\centering\arraybackslash}m{#1}}
\newcolumntype{L}[1]{>{\raggedright\arraybackslash}m{#1}}

\definecolor{lightGrey}{RGB}{240,240,240}
\definecolor{darkGrey}{RGB}{200,200,200}
\definecolor{Grey}{RGB}{80,80,80}
\definecolor{X}{RGB}{255,0,0}

\aptLtoX[graphic=no,type=html]{}{\titleformat{\subsubsection}[runin]
{\itshape}{\thesubsubsection}{0.5em}{\hspace{0.5em}}[. ]
\titleformat{\paragraph}[runin]
{}{\theparagraph}{0.5em}{\hspace{0.5em}}[. ]}

\AtBeginDocument{%
  }

\copyrightyear{2024}
\acmYear{2024}
\setcopyright{acmlicensed}
\acmConference[CHI '24]{Proceedings of the CHI Conference on Human Factors in Computing Systems}{May 11--16, 2024}{Honolulu, HI, USA}
\acmBooktitle{Proceedings of the CHI Conference on Human Factors in Computing Systems (CHI '24), May 11--16, 2024, Honolulu, HI, USA}
\acmDOI{10.1145/3613904.3642151}
\acmISBN{979-8-4007-0330-0/24/05}

\begin{document}


\title{"At the end of the day, I am accountable": Gig Workers' Self-Tracking for Multi-Dimensional Accountability Management}
\renewcommand{\shorttitle}{Gig Workers' Self-Tracking for Multi-Dimensional Accountability Management}

 \author{Rie Helene (Lindy) Hernandez}
  \affiliation{
   \institution{The Pennsylvania State University}
   \city{State College}
   \state{PA}
   \country{USA}
 }
 \email{lindyhernandez@psu.edu}
 \author{Qiurong Song}
  \affiliation{
   \institution{The Pennsylvania State University}
   \city{State College}
   \state{PA}
   \country{USA}
 }
 \email{qiurongsong@psu.edu}
 \author{Yubo Kou}
  \affiliation{
   \institution{The Pennsylvania State University}
   \city{State College}
   \state{PA}
   \country{USA}
 }
 \email{yubokou@psu.edu}
 \author{Xinning Gui}
 \affiliation{
   \institution{The Pennsylvania State University}
   \city{State College}
   \state{PA}
   \country{USA}
 }
 \email{xinninggui@psu.edu}


\renewcommand{\shortauthors}{Hernandez et al.}

\begin{abstract}
Tracking is inherent in and central to the gig economy. Platforms track gig workers' performance through metrics such as acceptance rate and punctuality, while gig workers themselves engage in self-tracking. Although prior research has extensively examined how gig platforms track workers through metrics – with some studies briefly acknowledging the phenomenon of self-tracking among workers – there is a dearth of studies that explore how and why gig workers track themselves. To address this, we conducted 25 semi-structured interviews, revealing how gig workers self-track to manage accountabilities to themselves and external entities across three identities: the holistic self, the entrepreneurial self, and the platformized self. We connect our findings to neoliberalism, through which we contextualize gig workers' self-accountability and the invisible labor of self-tracking. We further discuss how self-tracking mitigates information and power asymmetries in gig work and offer design implications to support gig workers’ multi-dimensional self-tracking.
\end{abstract} 



\maketitle

\section{Introduction}
Tracking workers’ behavior and performance has been a longstanding practice in various industries and economies \cite{Ravid_etal_2020, Stanton_2000}. This practice has evolved with the rise of platformization where industries adopt digital platforms as core infrastructure for business and services, accelerating labor and economic shifts \cite{Casilli_Posada_2019}. This transformation is especially pronounced within the gig economy, where platforms and algorithms play a central role in managing and monitoring workers \cite{Casilli_Posada_2019, Duggan_etal_2019}. Gig platforms leverage technology to track various aspects of gig workers' activities, including location and performance metrics (e.g., acceptance rate, cancellation rate, customer ratings) for control, safety, and customer satisfaction \cite{Mosseri_2022, Verelst_etal_2022}. Researchers have extensively studied these algorithms and metrics due to their role in surveilling and monitoring the worker \cite{Chan_2022, Lee_etal_2015}, aligning with broader studies on workplace tracking that delve into the implications of surveillance \cite{Bakewell_etal_2018, Pritchard_etal_2015}.

There is a distinction between the external tracking done by gig platforms and self-tracking for empowerment and self-knowledge. While there is an abundance of research on platform metrics, only some studies touch on gig workers’ self-tracking practices outside of gig platforms. For instance, gig workers have been observed to self-track as a form of protection from platforms and customers \cite{Mosseri_2022, Sannon_etal_2022}. Additionally, gig workers self-track income, expenses, and mileage to fulfill tax obligations given their classification as independent contractors which requires them to calculate and file income and self-employment taxes \cite{Oei_Ring_2017a, Thomas_2018}. Researchers have recognized gig workers' self-tracking practices and what necessitates them within these contexts, however, there are no focused studies that delve into understanding why they self-track, what they track, and how they use self-tracked data on a broader scale.

Within human-computer interaction, self-tracking is also commonly studied in the personal informatics (PI) field. The areas of workplace \cite{Lushnikova_etal_2022, Chung_etal_2017} and productivity tracking \cite{Kim_etal_2019, Guillou_etal_2020} within PI share some similarities with gig workers as they are concerned with labor, but largely focus on the traditional workforce. Additionally, some HCI researchers have explored how to enhance gig workers’ reflecting and acting on their own data \cite{You_etal_2021, Zhang_etal_2023}, based on the stage-based model of PI \cite{Li_etal_2010}. While PI has begun to be studied within the context of gig work, researchers have yet to comprehensively understand gig workers' behavior and activity of self-tracking.

Given these identified research gaps, we aimed to answer the following research questions: 1) Why do gig workers self-track? and 2) How do gig workers self-track? (i.e., what tools do they use, what data do they collect, \textcolor{black}{how do they use their data}). To answer these, we conducted 25 semi-structured interviews with gig drivers (i.e., rideshare, delivery, courier) from a diverse set of gig platforms who use a variety of self-tracking tools in their work lives. Applying an inductive thematic analysis approach to our data, it became apparent how gig workers use self-tracking to uphold accountabilities they hold across three distinct identities: personal accountabilities of the holistic self, fiscal and resource accountabilities of the entrepreneurial self, and performance accountabilities of the platformized self. Based on these findings, we discuss the self-accountability gig workers demonstrate and how self-tracking is a form of invisible labor. We further discuss self-tracking’s role in mitigating information and power asymmetries in gig work and provide design implications for multi-dimensional self-tracking.

With this study, we make several contributions to the field of HCI. First, we provide empirical insights into how gig workers utilize self-tracking to effectively manage their multi-dimensional accountabilities. This sheds light on the intricacies of the gig economy, which holds relevance for both academic inquiry and practical applications. Second, our research enriches the ongoing discourse concerning the prevalent information and power asymmetries within the gig economy. Third, we present design implications aimed at facilitating gig workers' intricate task of managing multi-dimensional accountability through self-tracking. These contributions collectively enhance our understanding of the gig economy's digital landscape and its implications for workers. 
\section{Related Work}
\textcolor{black}{The gig economy refers to a labor market characterized by short-term, flexible jobs, often facilitated through digital platforms or apps. In this model, individuals, referred to as gig workers, operate as independent contractors with increased autonomy, taking on tasks or projects on a temporary or part-time basis \cite{Manyika_etal_2016}. In a recent survey, McKinsey found that 36\% of employed U.S. respondents (around 58 million individuals) identify as independent workers, growing from 27\% in 2016 \cite{Dua_etal_2022}.  Examples of gig work include rideshare (e.g., Uber), food delivery (e.g., Doordash), courier (e.g., Roadie), freelance work (e.g., Fiverr,), and crowdwork (e.g., Amazon Mechanical Turk). Different services exhibit distinct characteristics, encompassing variations in service models, customer interaction, location, and complexity \cite{Duggan_etal_2019, Vallas_Schor_2020, Hsieh_etal_2023}.}

\aptLtoX[graphic=no,type=html]{
\begin{table*}[b]
 \centering
 \caption{Gig Platform Metrics. Where DR = Driver Rating; AR = Acceptance Rate; CR = Cancellation Rate; CSR = Completion/Success Rate; PR = Punctuality Rate; DRC = Delivered \& Received Count; SCR = Schedule Commitment Rate; SA = Shopping Accuracy; SPI = Seconds per Item.}~\label{tab:platform-metrics}
 \begin{tabular}{|l|l|l|l|l|l|l|l|l|l|l|l} 
 
\hline
\rowcolor[HTML]{e6e6e6}
{\textbf{Platform}}
 & { \textbf{Category}}
 & { \textbf{DR}}
 & { \textbf{AR}}
 & { \textbf{CR}}
 & { \textbf{CSR}}
 & { \textbf{PR}}
 & { \textbf{DRC}}
 & { \textbf{SCR}}
 & { \textbf{SA}}
 & { \textbf{SPI}}
  \\
 \hline
 
Lyft & Rideshare & \checkmark & \checkmark & \checkmark &&&&&& \\ \hline

Uber & Rideshare & \checkmark & \checkmark & \checkmark &&&&&& \\ \hline

Amazon Flex & Delivery &&& \checkmark & \checkmark & \checkmark & \checkmark &&& \\ \hline

DoorDash & Delivery & \checkmark & \checkmark && \checkmark & \checkmark &&&& \\ \hline

Grubhub & Delivery && \checkmark & \checkmark & \checkmark & \checkmark && \checkmark && \\ \hline

Instacart & Delivery & \checkmark && \checkmark &&&&& \checkmark & \checkmark \\ \hline

Shipt & Delivery & \checkmark &&& \checkmark & \checkmark &&&& \\ \hline

Uber Eats & Delivery & \checkmark & \checkmark & \checkmark &&&&&& \\ \hline

Walmart Spark & Delivery & \checkmark & \checkmark & \checkmark && \checkmark &&&& \\ \hline

 \end{tabular}
\end{table*}}{
\begin{table*}[b]
 \sffamily\smaller
 \def\arraystretch{1.3}
 \centering
 \caption{Gig Platform Metrics. Where DR = Driver Rating; AR = Acceptance Rate; CR = Cancellation Rate; CSR = Completion/Success Rate; PR = Punctuality Rate; DRC = Delivered \& Received Count; SCR = Schedule Commitment Rate; SA = Shopping Accuracy; SPI = Seconds per Item.}~\label{tab:platform-metrics}
\scalebox{0.88}{
 \begin{tabular}{!{\color{black}\vrule} L{0.145\textwidth} | L{0.105\textwidth} | L{0.055\textwidth} | L{0.055\textwidth} | L{0.055\textwidth} | L{0.055\textwidth} | L{0.055\textwidth} | L{0.055\textwidth} | L{0.055\textwidth} | L{0.055\textwidth} | L{0.055\textwidth} | L{0.055\textwidth} !{\color{black}\vrule}} 
 
\arrayrulecolor{black}\hline
\rowcolor{lightGrey}
{\smaller \textbf{Platform}}
 & {\smaller \textbf{Category}}
 & {\smaller \textbf{DR}}
 & {\smaller \textbf{AR}}
 & {\smaller \textbf{CR}}
 & {\smaller \textbf{CSR}}
 & {\smaller \textbf{PR}}
 & {\smaller \textbf{DRC}}
 & {\smaller \textbf{SCR}}
 & {\smaller \textbf{SA}}
 & {\smaller \textbf{SPI}}
  \\
 \hline
 
Lyft & Rideshare & \checkmark & \checkmark & \checkmark &&&&&& \\ \hline

Uber & Rideshare & \checkmark & \checkmark & \checkmark &&&&&& \\ \hline

Amazon Flex & Delivery &&& \checkmark & \checkmark & \checkmark & \checkmark &&& \\ \hline

DoorDash & Delivery & \checkmark & \checkmark && \checkmark & \checkmark &&&& \\ \hline

Grubhub & Delivery && \checkmark & \checkmark & \checkmark & \checkmark && \checkmark && \\ \hline

Instacart & Delivery & \checkmark && \checkmark &&&&& \checkmark & \checkmark \\ \hline

Shipt & Delivery & \checkmark &&& \checkmark & \checkmark &&&& \\ \hline

Uber Eats & Delivery & \checkmark & \checkmark & \checkmark &&&&&& \\ \hline

Walmart Spark & Delivery & \checkmark & \checkmark & \checkmark && \checkmark &&&& \\ \hline

\arrayrulecolor{black}\hline
 \end{tabular}
}
\end{table*}}
\aptLtoX[graphic=no,type=html]{
\begin{table*}[b]
 \centering
 \caption{Gig Platform In-App Tracked Information. Each platform will display tracked data differently, whether it's on a weekly, daily, or per trip basis.}~\label{tab:tracked-info}
 \begin{tabular}{|l|l|l|l|l|l|} 
 
\hline
\rowcolor[HTML]{e6e6e6}
{ \textbf{Platform}}
 & { \textbf{Category}}
 & { \textbf{Mileage}}
 & { \textbf{Income}}
 & { \textbf{Working Time}}
 & {\textbf{Number of Gigs}}
 \\ 
 \hline
 
Lyft & Rideshare & \checkmark & \checkmark & \checkmark & \checkmark  \\ \hline

Uber & Rideshare & \checkmark & \checkmark & \checkmark & \checkmark  \\ \hline

Amazon Flex & Delivery && \checkmark & \checkmark & \\ \hline

DoorDash & Delivery & \checkmark & \checkmark & \checkmark & \checkmark  \\ \hline

Grubhub & Delivery & \checkmark & \checkmark && \\ \hline

Instacart & Delivery & \checkmark & \checkmark & \checkmark & \checkmark  \\ \hline

Shipt & Delivery && \checkmark && \checkmark  \\ \hline

Uber Eats & Delivery & \checkmark & \checkmark & \checkmark & \checkmark  \\ \hline

Walmart Spark & Delivery & \checkmark & \checkmark && \checkmark  \\ \hline

 \end{tabular}
\end{table*}}{
\begin{table*}[b]
 \sffamily\smaller
 \def\arraystretch{1.3}
 \centering
 \caption{Gig Platform In-App Tracked Information. Each platform will display tracked data differently, whether it's on a weekly, daily, or per trip basis.}~\label{tab:tracked-info}
\scalebox{0.88}{
 \begin{tabular}{!{\color{black}\vrule} L{0.145\textwidth} | L{0.105\textwidth} | L{0.095\textwidth} | L{0.095\textwidth} | L{0.095\textwidth} | L{0.095\textwidth} !{\color{black}\vrule}} 
 
\arrayrulecolor{black}\hline
\rowcolor{lightGrey}
{\smaller \textbf{Platform}}
 & {\smaller \textbf{Category}}
 & {\smaller \textbf{Mileage}}
 & {\smaller \textbf{Income}}
 & {\smaller \textbf{Working Time}}
 & {\smaller \textbf{Number of Gigs}}
 \\ 
 \hline
 
Lyft & Rideshare & \checkmark & \checkmark & \checkmark & \checkmark  \\ \hline

Uber & Rideshare & \checkmark & \checkmark & \checkmark & \checkmark  \\ \hline

Amazon Flex & Delivery && \checkmark & \checkmark & \\ \hline

DoorDash & Delivery & \checkmark & \checkmark & \checkmark & \checkmark  \\ \hline

Grubhub & Delivery & \checkmark & \checkmark && \\ \hline

Instacart & Delivery & \checkmark & \checkmark & \checkmark & \checkmark  \\ \hline

Shipt & Delivery && \checkmark && \checkmark  \\ \hline

Uber Eats & Delivery & \checkmark & \checkmark & \checkmark & \checkmark  \\ \hline

Walmart Spark & Delivery & \checkmark & \checkmark && \checkmark  \\ \hline

\arrayrulecolor{black}\hline
 \end{tabular}
}
\end{table*}}

\textcolor{black}{As independent contractors, gig workers shoulder a unique form of accountability as they self-manage the intricacies of their roles, marked by a notable degree of autonomy and responsibility as a result of platformization and algorithmic management. They must manage various aspects of their work lives from before, during, and after working to maintain an advantage and stay profitable on that platform \cite{Alvarez_etal_2023, Hsieh_etal_2022}. Existing studies on self-management have predominantly focused on business leaders \cite{Ghanem_Castelli_2019, Dose_Klimoski_1995}, who, similar to independent contractors, operate with a high degree of autonomy and are responsible for managing their work without immediate oversight. In these studies, the concept of self-management is closely intertwined with self-accountability as it acts as an internal control system in the absence of being answerable to another party \cite{Ghanem_Castelli_2019, Dose_Klimoski_1995}. Self-management, as defined in studies with independent contractors and business leaders, involves exerting heightened personal control over one's career in the absence of direct supervision. It includes a broad set of activities (i.e., defining working hours, managing finances, and setting goals) that allow individuals to enhance productivity, achieve objectives, and make work-related progress. While self-management typically employs various techniques to reach diverse goals, our focus centers on exploring how gig workers self-track to manage their unique needs and situations.}

\textcolor{black}{Gig workers in the U.S. must track information about themselves (e.g., income and mileage) for tax reporting purposes \cite{Helling_2019}. To monitor these essential data points and other kinds of information, gig workers have access to various tools including gig platforms, third-party tools, and manual tracking methods. Gig platforms provide in-app tracking features such as platform metrics (Table \ref{tab:platform-metrics}) or activity information (Table \ref{tab:tracked-info}). Platform metrics are measurements used to assess performance, while activity information consists of information that are not direct performance indicators. The tables below present tracked metrics and information on various gig platform driver apps, organized based on their functions. There are also several third-party applications (e.g., Gridwise, Stride) that cater to gig workers. These typically include mileage and expense tracking for tax needs and can include features like smart tools to maximize earnings or information on high-demand areas. Workers can also manually track their activity using pen and paper, or digital spreadsheets like Google Sheets or Microsoft Excel.}

\textcolor{black}{Below, we examine existing literature on accountability and self-tracking practices among gig workers. We first review accountability literature and connect this to existing discussions on gig worker accountability. We then examine personal informatics literature and connect this to how self-tracking has been observed or studied amongst gig workers.}

\subsection{Accountability and Gig Work}
Accountability and responsibility are often conflated with one another, but there lie subtle differences in how each is felt and assigned. Responsibility implies self-control and obligation, while accountability involves being answerable to another party (i.e., a coworker, supervisor, or customer) or to oneself \cite{Dose_Klimoski_1995, Pagan_etal_2022}. Along with being answerable is the expectation that an “evaluation may occur” and that an “explanation [by the one held accountable] is required” \cite{Hall_etal_2017}. While the two can be used interchangeably, responsibility is the broader of the two as it encompasses a range of obligations, duties, and tasks that may include an individual’s accountabilities. We use accountability for this study as it is a more relational concept that captures the intricacies of various relationships, expectations, and influences that shape gig workers’ behavior and decisions. \textcolor{black}{To understand this subject, we review literature that has explored the accountability of self-employed individuals, contractors, and gig workers.}

\subsubsection{\textcolor{black}{Self-Employed and Contractor Accountability}}
Studies on self-employed individuals have specifically looked into their financial accountabilities given the increased responsibilities and enhanced complexities surrounding their self-employment \cite{Boden_1999, Ritchie_Richardson_2000}. Here, researchers explore the increased accountabilities self-employed individuals have surrounding managing their finances and taxes as they are accountable to systems in place – rather than individuals – such as tax or child-care organizations \cite{Boden_1999}. For contractors in the government \cite{Blomqvist_Winblad_2022, Delfino_2022} and military \cite{Hackman_2007, Mehra_2009, Mulgan_2006}, researchers mainly explore accountability in adverse incidents: who is and should be legally held accountable, and whether that ownership makes sense considering the contractor-employer relationship. There are similar discussions surrounding accountability in accidents involving gig workers. There have been cases leading to injury or death where gig platforms avoided accountability as they argue their position as a matching or networking service instead of a rideshare or delivery company \cite{Codagnone_etal_2016}. Drivers found themselves in a gray area when accidents occurred while logged on and looking for another gig, but not while fulfilling a delivery or ride service \cite{Pfeffer-Gillett_2016}. These instances highlight gig workers’ independent contractor classification, how platform companies distance themselves from the worker to maintain this classification \cite{Rauch_2021}, and how these companies exploit employee classification systems to reap “rewards without risk and responsibility” \cite{Woodcock_Graham_2020}.

\subsubsection{\textcolor{black}{Gig Worker Accountability}}
Explorations into gig workers’ accountabilities also consider how accountability is shifted between the worker and platform due to workers’ classification as independent contractors – particularly what the platform will take accountability for and what they leave to the worker. Researchers have observed how gig workers are accountable for maintaining separate records of their activity that are also automatically and algorithmically evaluated by the platform \cite{Mosseri_2022, Sannon_etal_2022}. As a result, they must manage their standing with “reputation auditing” where they rectify platform records to more accurately reflect their experiences \cite{Mosseri_2022}. Workers do this by maintaining records of their activity which they use to support claims, thus redirecting the accountability they previously held to the platform or customer \cite{Rosenblat_Stark_2016, Sannon_etal_2022}. These reports signify the precarious relationship between contractor and employer as they serve as the focal point of accountability negotiation between the two parties.

In summary, accountability researchers have primarily explored how contractors and self-employed workers manage their unique accountability dynamics with companies or not having anyone to directly report to. While the subjects of these studies draw similarities to gig workers, the elements of their employment still differ, particularly when considering the ambiguous boundary they walk between contractor and employee \cite{Rauch_2021}. Studies exploring gig worker accountability highlight how their classification as independent contractors affects how accountability is managed and delegated between themselves and the platforms they serve. While these studies delve into the construction of accountability in these relationships, they don’t explore how gig workers internally manage accountability. This study addresses this gap by reporting on how gig workers self-track to oversee their accountabilities.

\subsection{Personal Informatics and Gig Work}
Personal informatics (PI) is “a school of thought which aims to use technology for acquiring and collecting data on different aspects of the daily lives of people” \cite{Rapp_Cena_2014}. PI systems are those that “help people collect personally relevant information for self-reflection and gaining self-knowledge” \cite{Li_etal_2010}. Research in PI has explored self-tracking across various populations such as Quantified Selfers \cite{Choe_etal_2014} and knowledge workers \cite{Kim_etal_2019}, as well as diverse domains like menstruation \cite{Epstein_etal_2017}, physical activity \cite{Tong_etal_2016}, and finance \cite{Kaye_etal_2014}. Notably, the field has heavily focused on health and wellbeing, data collection and reflection, and behavior change motivations \cite{Epstein_etal_2020}. 

\subsubsection{\textcolor{black}{Workplace and Productivity Tracking}}
While gig workers’ self-tracking activities have only sparsely been studied by researchers, two more commonly explored areas of workplace and productivity tracking draw some similarities to gig work. Workplace and productivity tracking both explore how tracking can be used to enhance activity, performance, and output, with workplace tracking being broader as it also encompasses work-related productivity tracking. Workplace tracking involves tracking to support, raise awareness of, or manage individuals in the workplace. Research has highlighted its benefits in enhancing collaboration among colleagues \cite{Lushnikova_etal_2022}, improving workplace health and wellbeing \cite{Chung_etal_2017}, and increasing productivity \cite{Kim_etal_2019}. Closely related is productivity tracking which has been understood in both the workplace context and on a more individual level. Researchers have explored ways to improve \cite{Kim_etal_2019} or quantify \cite{Kim_etal_2016} productivity. Productivity researchers have looked into how tracking productivity can enhance wellbeing as workers balance and reflect on productive time alongside breaks \cite{DiLascio_etal_2020, Guillou_etal_2020}, leading to increased awareness and perception of their work \cite{Guillou_etal_2020, Kim_etal_2019}. However, while researchers have highlighted the utility of both workplace and productivity tracking, they also note negative side effects such as issues of surveillance \cite{Bakewell_etal_2018, Pritchard_etal_2015}, heightened stress \cite{Moore_2019}, and pressure to track \cite{Tackx_etal_2021, Heikkilä_etal_2018}.

Workplace and productivity tracking studies are somewhat related to gig workers as they center on labor, however, the two primarily focus on traditional workplaces where employees answer to a supervisor or work collaboratively with others. The study subjects are often information and knowledge workers such as software developers, designers, and researchers \cite{Epstein_etal_2016, Kim_etal_2019, Luo_etal_2018, Meyer_etal_2014}. \textcolor{black}{Tracking in these contexts differs from that of gig workers as they lack direct supervisors, necessitating increased self-management and accountability.} Gig workers’ self-tracking use and experiences remain understudied, but some HCI researchers have begun investigating ways to enhance gig workers’ reflecting and acting on their data. One study explored using social sensing (a collaborative effort with drivers’ partners to understand and make decisions based on their data) to help drivers understand their health and achieve work-life balance \cite{You_etal_2021}. Another explored the challenges surrounding algorithmic management by focusing on how the design and presentation of data influence the wellbeing and work experience of gig workers \cite{Zhang_etal_2023}. The researchers found that their designs served as boundary objects in AI, helping participants gain insights about their work activity and experiences.

\subsubsection{\textcolor{black}{Gig Worker Self-Tracking}}
\textcolor{black}{Some researchers have noted gig workers' self-tracking activity concerning their tax responsibilities \cite{Oei_Ring_2017b, Oei_Ring_2017a} and self-protection \cite{Mosseri_2022, Sannon_etal_2022, Cameron_2022}; however, there has not been an in-depth empirical study of these tracking behaviors.} Gig workers’ self-employed status increases their tax responsibilities compared to employees of a company, placing additional administrative burdens on their shoulders \cite{Thomas_2018}. These include self-reporting income and maintaining thorough records for potential IRS audits \cite{Oei_Ring_2017a}. Failure to maintain records of their income, expenses, or mileage may result in tax challenges \cite{Oei_Ring_2017b, Thomas_2018}. Furthermore, while some gig platforms track workers’ income and mileage, the reported mileage is often an incomplete account of the worker’s actual business mileage \cite{Oei_Ring_2017a, Oei_Ring_2017b}. As a result of gig workers’ unique tax circumstances, researchers in this field have identified the need for gig workers to engage in self-tracking as they relate to their tax requirements.

Self-tracking has also been observed as a response to the algorithmic management and platform surveillance that gig workers experience. Platform algorithms compare worker data against company-specific metrics as indicators of performance \cite{vanDoorn_Badger_2020, Verelst_etal_2022}, leading to workers constantly monitoring their metrics \textcolor{black}{to assess their performance \cite{Cameron_2022}} and avoid consequences such as fewer job offers or deactivation \cite{Kirven_2018, Manokha_2020}. As such, this algorithmic management is often referred to as a form of workplace surveillance as platforms seek to control workers’ performance and behavior on the job \cite{Moore_Joyce_2020, Newlands_2022}. Workplace tracking researchers have similarly raised concerns surrounding increased surveillance in the workplace, drawing comparisons to Foucault’s panopticon \cite{Bakewell_etal_2018, Pritchard_etal_2015}. Platform surveillance prompts workers to engage in self-protective measures such as \textcolor{black}{tracking conversations with support to ensure issues are resolved \cite{Cameron_2022}}, record-keeping as a way to protect themselves in disputes \cite{Sannon_etal_2022}, and “reputation auditing” with their tracked data as a way of resolving poor metrics \cite{Mosseri_2022}. This body of research has observed the use of self-tracking and the motivation behind it as a form of self-protection, however, it lacks an in-depth exploration of self-tracking behavior.

In summary, despite research in workplace and productivity tracking sharing some similarities with gig work as they all center on labor, studies here predominantly focus on knowledge workers in traditional settings. \textcolor{black}{This differs significantly from non-traditional work arrangements in the gig economy where workers lack direct supervision and tracking is more closely tied to financial and legal implications, such as their heightened tax accountabilities.} Furthermore, while researchers have begun to recognize the self-tracking needs of gig workers, there is a lack of grounding in empirical research that focuses on why and how they self-track. Existing gig worker studies show that self-tracking is used in various dimensions of a worker’s activity, prompting an investigation into how self-tracking among gig workers might result in unique practices and experiences. Our study provides an in-depth understanding of how gig workers use self-tracking to maintain accountability given their categorization as self-employed, independently contracted workers. 
\section{Methods}
We conducted semi-structured interviews with 25 gig drivers to understand their use of self-tracking tools and how these tools support their work, then used an inductive thematic analysis \cite{Braun_Clarke_2019, Clarke_Braun_2021} to analyze the data.

\aptLtoX[graphic=no,type=html]{
\begin{table*}[b]
 \centering
 \caption{Participant Demographics and Tracking Tools Used.}~\label{tab:participants}
 \begin{tabular}{|l|l|l|l|l|l|} 
 
\hline
\rowcolor[HTML]{e6e6e6}
{ \textbf{P\#}}
 & { \textbf{Age}}
 & { \textbf{Race}}
 & { \textbf{Gender}}
 & { \textbf{Gig Platform/s}}
 & { \textbf{Tracking Tool/s}}
 \\ 
 \hline
 
1 & 24 & Black or African American & Male & DoorDash & Triplog, Spreadsheet  \\ \hline

2 & 30 & Black or African American & Female & Uber & Gridwise, Spreadsheet  \\ \hline

3 & 45 & White & Male & Grubhub, Uber Eats & Pen and paper  \\ \hline

4 & 25 & Black or African American & Female & DoorDash, Grubhub & Pen and paper, Microsoft Excel  \\ \hline

5 & 27 & Asian & Nonbinary & Roadie & Microsoft Excel  \\ \hline

6 & - & White & Female & DoorDash, Walmart Spark & Stride, Excel, GasBuddy  \\ \hline

7 & 44 & White & Male & Amazon Flex & MileIQ, Microsoft Excel  \\ \hline

8 & 31 & White & - & DoorDash, GoPuff & SportsTracker, Waze, GasBuddy, GPS Camera, Notes app  \\ \hline

9 & 62 & White & - & Amazon Flex & Pen and paper  \\ \hline

10 & 58 & White, Hispanic/ Spanish/ Latino & Male & DoorDash, Uber Eats, Walmart Spark, Roadie, Shipt & Gridwise, Spreadsheet  \\ \hline

11 & 30 & Black or African American &    Male & Uber, Lyft, DoorDash, Instacart & Everlance, Fuelio, Gridwise  \\ \hline

12 & 30 & Black or African American & Male & Lyft & Lyft app  \\ \hline

13 & 40 & American Indian or Alaska Native & Nonbinary & Uber Eats & Uber Eats app, Pen and paper  \\ \hline

14 & 32 & White & Male & Amazon Flex, Instacart, DoorDash, Grubhub, Favor & Google Sheets  \\ \hline

15 & 36 & White & Male & DoorDash, Grubhub, Uber Eats, Roadie, Shipt & MileIQ, Google Sheets  \\ \hline

16 & 27 & Black or African American & Male & Uber & Notes app  \\ \hline

17 & 28 & White & Female & Uber & Uber app  \\ \hline

18 & 25 & Black or African American & Female & Uber Eats & Uber Eats app, Google Maps, Apple Maps  \\ \hline

19 & 27 & Black or African American & Male & Uber, Uber Eats & Everlance, Gridwise, Fuelio  \\ \hline

20 & 30 & American Indian or Alaska Native, Black or African American & Male & Uber & Gridwise  \\ \hline

21 & 36 & White & Male & DoorDash & Everlance  \\ \hline

22 & 31 & White & Male & Uber & Everlance, Notes app  \\ \hline

23 & 25 & Black or African American & Male & Uber, Lyft & Traqq, Microsoft Excel  \\ \hline

24 & 25 & Black or African American & Female & DoorDash & Para  \\ \hline

25 & 30 & Asian & Female & Uber Eats, Shipt & Everlance, Mint  \\ \hline

 \end{tabular}

\end{table*} }{
\begin{table*}[b]
 \sffamily\smaller
 \def\arraystretch{1.3}
 \centering
 \caption{Participant Demographics and Tracking Tools Used.}~\label{tab:participants}
\scalebox{0.88}{
 \begin{tabular}{!{\color{black}\vrule} L{0.055\textwidth} | L{0.055\textwidth} | L{0.225\textwidth} | L{0.095\textwidth} | L{0.24\textwidth} | L{0.24\textwidth} !{\color{black}\vrule}} 
 
\arrayrulecolor{black}\hline
\rowcolor{lightGrey}
{\smaller \textbf{P\#}}
 & {\smaller \textbf{Age}}
 & {\smaller \textbf{Race}}
 & {\smaller \textbf{Gender}}
 & {\smaller \textbf{Gig Platform/s}}
 & {\smaller \textbf{Tracking Tool/s}}
 \\ 
 \hline
 
1 & 24 & Black or African American & Male & DoorDash & Triplog, Spreadsheet  \\ \hline

2 & 30 & Black or African American & Female & Uber & Gridwise, Spreadsheet  \\ \hline

3 & 45 & White & Male & Grubhub, Uber Eats & Pen and paper  \\ \hline

4 & 25 & Black or African American & Female & DoorDash, Grubhub & Pen and paper, Microsoft Excel  \\ \hline

5 & 27 & Asian & Nonbinary & Roadie & Microsoft Excel  \\ \hline

6 & - & White & Female & DoorDash, Walmart Spark & Stride, Excel, GasBuddy  \\ \hline

7 & 44 & White & Male & Amazon Flex & MileIQ, Microsoft Excel  \\ \hline

8 & 31 & White & - & DoorDash, GoPuff & SportsTracker, Waze, GasBuddy, GPS Camera, Notes app  \\ \hline

9 & 62 & White & - & Amazon Flex & Pen and paper  \\ \hline

10 & 58 & White, Hispanic/ Spanish/ Latino & Male & DoorDash, Uber Eats, Walmart Spark, Roadie, Shipt & Gridwise, Spreadsheet  \\ \hline

11 & 30 & Black or African American &    Male & Uber, Lyft, DoorDash, Instacart & Everlance, Fuelio, Gridwise  \\ \hline

12 & 30 & Black or African American & Male & Lyft & Lyft app  \\ \hline

13 & 40 & American Indian or Alaska Native & Nonbinary & Uber Eats & Uber Eats app, Pen and paper  \\ \hline

14 & 32 & White & Male & Amazon Flex, Instacart, DoorDash, Grubhub, Favor & Google Sheets  \\ \hline

15 & 36 & White & Male & DoorDash, Grubhub, Uber Eats, Roadie, Shipt & MileIQ, Google Sheets  \\ \hline

16 & 27 & Black or African American & Male & Uber & Notes app  \\ \hline

17 & 28 & White & Female & Uber & Uber app  \\ \hline

18 & 25 & Black or African American & Female & Uber Eats & Uber Eats app, Google Maps, Apple Maps  \\ \hline

19 & 27 & Black or African American & Male & Uber, Uber Eats & Everlance, Gridwise, Fuelio  \\ \hline

20 & 30 & American Indian or Alaska Native, Black or African American & Male & Uber & Gridwise  \\ \hline

21 & 36 & White & Male & DoorDash & Everlance  \\ \hline

22 & 31 & White & Male & Uber & Everlance, Notes app  \\ \hline

23 & 25 & Black or African American & Male & Uber, Lyft & Traqq, Microsoft Excel  \\ \hline

24 & 25 & Black or African American & Female & DoorDash & Para  \\ \hline

25 & 30 & Asian & Female & Uber Eats, Shipt & Everlance, Mint  \\ \hline

\arrayrulecolor{black}\hline
 \end{tabular}
}
\end{table*}}

\subsection{Participant Recruitment}
We focused on recruiting gig drivers (i.e., rideshare, food delivery, and courier drivers) due to their unique tracking behaviors where they track not only time and finances but also mileage for tax deductions \cite{Thayer_2020}. This data dimension offers insights into their self-tracking practices and data interpretation, distinct from other kinds of gig work. Microtaskers (e.g., Task Rabbit, Handy) were excluded as they are less common than rideshare and delivery roles available across most U.S. cities. Additionally, rideshare and delivery drivers do similar enough jobs that there are overlaps in those groups.

Recruitment was done both online and offline. For online recruitment, we planned to post solely on gig work-related forums (subreddits) on Reddit that allowed calls for research participants but found that several subreddits prohibited posting about research studies. Therefore, we expanded the search to include non-gig work-targeted subreddits and were able to post (with permission) on a total of eight subreddits (r/GigWork, r/amazonflexdrivers, r/UberEATS, r/Grubhubdrivers, r/roadie, r/SampleSize, and a subreddit for the authors’ university). Despite the limitation, we were still able to recruit a diverse set of participants due to the phenomenon of “multi-apping”, where gig workers work with a variety of platforms at a time instead of being constrained to just one \cite{Goods_etal_2019}. For offline recruitment, we created and distributed flyers to various local establishments (e.g., restaurants, cafes, community centers, school buildings) to reach gig workers in our community. 

All recruitment messages included information on the study goal, expected interview length, compensation, inclusion criteria, and a link to the screener survey. This survey collected preliminary data to select a diverse group of participants based on demographics (e.g., How old are you?, To which gender identity do you most identify?), gig work specifics (e.g., What specific gig platform/s do you currently work on?), and self-tracking tool use (e.g., What apps or tools do you use for tracking related to your job?). Of the 380 individuals who completed the screener survey, 274 met the inclusion criteria: being in the U.S., aged 18 or older, working as a gig driver (e.g., rideshare, courier, delivery), and self-tracking for gig work. We used purposeful sampling techniques \cite{Palinkas_etal_2015} to select a diverse sample in terms of their background (i.e., race, gender, age) and their gig work activity (i.e., platforms they worked on, tracking applications they used). 

We interviewed 25 participants whose demographics and gig work details can be found in Table \ref{tab:participants}. Ages ranged from 24 to 62 years (median = 30). Of the participants, 14 identified as male, 7 identified as female, 2 identified as non-binary, and 2 did not disclose. For types of gig work, 17 participants did some sort of delivery (i.e., food delivery or courier service), 6 did rideshare, and 2 did both delivery and rideshare. The difference between rideshare and delivery drivers can be explained by several participants stopping rideshare and moving to delivery due to the COVID-19 pandemic, aligning with gig worker activity across the U.S. \cite{Regan_Christie_2022}. Participants used various tracking methods including automatic tracking applications, pen and paper, spreadsheets, and gig platform-based tracking features.

\subsection{Data Collection}
Interview questions covered the information they paid attention to and tracked (e.g., What information do you find useful to know to perform your job activities?, What information about you do the gig platforms track and present to you?), their choice of self-tracking tools (e.g., What made you select your tracking app?, What are your criteria for a tracking tool?), their use of self-tracking tools and self-tracked data (e.g., Why do you track?, What do you learn by looking at your data?), and their experiences with work-related tracking (e.g., How do you feel about tracking your data for work?, What challenges do you encounter while self-tracking for work?). Probing questions were dynamically created when needed to gather additional insights. After the initial ten interviews, the interview protocol was refined. The changes involved adding common follow-up questions (e.g., ‘How often do you look at this information?’ in regards to platform metrics), revising questions for clarity (e.g., ‘How do you make sense of data from different tracking sources?’ turned to ‘How do you evaluate or integrate data you have from different tracking apps?’), and eliminating questions with redundant answers (e.g., ‘Do you need to distinguish between data from different gig platforms?’ and ‘Do you have different tracking needs based on the different platforms?’).

Interviews were conducted in English, video and audio recorded, and transcribed verbatim, with recordings and transcriptions stored in a password-protected database. Participants provided verbal consent to being recorded, receiving detailed information on the study, their rights as volunteers, and data usage. They were also given the opportunity to ask questions about the study. Interviews lasted approximately 30-90 minutes (median = 47 minutes, average = 49.76 minutes). Upon completion, 24/25 participants received a \$30 Amazon gift card as compensation and one was compensated \$15 as they did not complete the interview. This participant was only able to partially complete the interview due to Internet connection issues, and we were unable to reschedule a follow-up interview. Nonetheless, they provided valuable insights into questions related to how they track, why they track, and how they use their tracked data for work-related activities. 

\subsection{Data Analysis}
We used reflexive thematic analysis to iteratively code data and generate themes with a bottom-up approach. Following Braun \& Clarke’s six phases \cite{Braun_Clarke_2019, Clarke_Braun_2021}, the first two authors familiarized themselves with the data by reviewing recordings and fixing transcriptions, then used Google Sheets to code by independently inspecting quotes and creating detailed descriptions of participants’ activities and experiences that were meaningful and relevant to the study. After coding, the two authors met to discuss codes by highlighting interesting points and reviewing similar or unclear codes. Following these meetings, the two authors met with the rest of the research team to discuss and refine codes as needed. The final set includes 557 unique codes such as “paying attention to platform metrics to avoid consequences (i.e., deactivation) that may happen if you reach a certain number,” “seeing self-tracking as something you have to do as a business,” and “tracking to get a picture of where they stand when calculating all income and expenses.” The first two authors then generated initial themes by grouping similar codes, an iterative process where the authors refined themes for cohesiveness and distinction. The full research team then reviewed candidate themes to ensure they made sense and were a representative analysis of the full dataset. Once the themes were set, the team proceeded with defining and naming themes to specify and differentiate them. Finally, the team began the writing process to weave together the analysis with data extracts to create a coherent account of the data. 
\section{Findings}
\textcolor{black}{Participants' self-tracking behaviors were closely linked to the accountabilities they held. Three identities crystallized as a result of observing and analyzing how participants navigated and expressed themselves concerning their roles and obligations, \textcolor{black}{shedding light on why and how gig workers engage in self-tracking}.} First is the holistic self where gig workers self-track to manage personal accountabilities and \textcolor{black}{empower themselves}. Second is the entrepreneurial self where gig workers consider fiscal and resource accountabilities as self-employed, independently contracted workers. Finally, we have the platformized self where workers manage performance accountabilities as they navigate relationships with the platforms and customers.

\subsection{Personal accountability to the holistic self}
Gig workers maintain personal accountabilities to both themselves and others. They are self-accountable for learning their capacity for work, making decisions, achieving work/life balance, and recognizing achievements. They are also accountable to their network, particularly when their work directly impacts those relationships. For instance, if a worker depends on their gig income to support their family, they are accountable for generating sufficient income for their family’s needs. These encapsulate the holistic selves that gig workers self-track to gather knowledge on and act in the interest of, enabling them to make informed decisions about when, how much, and which jobs to take. 

However, gig platforms do not provide adequate tools that can help workers track for their holistic selves. \textcolor{black}{For example, several participants, aligning with existing studies, reported incomplete mileage \cite{Oei_Ring_2017a, Oei_Ring_2017b} or incorrect income \cite{Cameron_2022} tracking by gig platforms. Additionally, our participants reported that platform-tracked information often centers around metrics or factors pertinent to the platform company (i.e., job-related metrics) which may not be relevant for workers' needs:} 

\begin{quote}
    “I would say [my tracking] has to do with personal stuff, but theirs has to do with business. So, [their tracking is] after probably their money, their sales, their business. For my [tracking], I’m improving myself, [working] on myself, [staying] out to make myself better than [...] the last day.” (P11)  
\end{quote}

Thus, participants used third-party tracking tools in addition to platform-provided information. With external tools, they can track personally relevant information that may not be tracked by gig platforms like time spent working, overall mileage, or personal experiences while working. \textcolor{black}{Additionally, by calculating, analyzing, and reviewing this data, they can learn what the job is like, determine if their jobs are worth continuing, know how much they're earning, monitor productivity, and manage work/life balance.} With this, workers \textcolor{black}{are empowered as they gain knowledge about themselves, reflect on their activity, and make informed decisions. Self-tracking allows them to cultivate their autonomy and flexibility as they can use their collected information to act in their own self-interest.}

\subsubsection{Generating self-knowledge}
Self-tracking grants gig workers a quantified and monitored understanding of their activity, providing them with valuable insights into the details and dynamics of their work lives which can be used to make professional and personal decisions. \textcolor{black}{Previous studies found that since workers navigate the gig landscape independently, they used online communities to learn about the job \cite{Yao_etal_2021, Holikatti_Jhaver_Kumar_2019}. Beyond this, we found that participants used self-tracking to understand gig work as it pertained to their unique experiences, particularly due to the market-dependent nature of gig work:}
\textcolor{black}{\begin{quote}
    "A lot of gig work [...] is market dependent. Los Angeles is going to [be] a lot different [from] where I am in Kentucky. So you can't compare your city to Los Angeles, they just have different food and demographics." (P6)
\end{quote}}

Each worker’s experience could significantly differ from another \textcolor{black}{due to differing local market conditions that impact job pricing and availability \cite{Casilli_Posada_2019, Duggan_etal_2019}. As a result,  there is limited information that they can exchange to learn about the gig work experience, implying that workers should explore more personalized methods to gain insights into working within their specific market.} For instance, self-tracking \textcolor{black}{mileage, working time, and income specifically helped P25 understand gig work as it pertained to their unique experiences:}

\begin{quote}
    “In the beginning with Shipt, when I was tracking, I also was just learning about gig work and how it worked, and really tracking how much I was making an hour, how much time I was working in coordination with how much I was making in a month. And so, I guess just like learning the capacity of how much I could make.” (P25) 
\end{quote}

Tracking also allowed them to learn more about the platforms they serve: “The more data you have points to the [...] things you can start tracking and seeing, and suddenly things come to light.” (P14). P14 shared how they were able to understand assignment algorithms by observing their tracked pay data where. From there, they noticed they were not paid consistent rates by Amazon: “If I never put together a spreadsheet for some of these base [pays] [...] I would have thought in my head it's just coincidence that some of these lower-paying shifts have been some of my worst shifts.” (P14). Having their tracked data available to them allowed them to see a pattern in what kinds of jobs were being sent their way.

Self-tracking \textcolor{black}{their activity }also gave workers an objective perspective of their activity to monitor their productivity, efficiency, and finances as they relate to their needs and goals:

\begin{quote}
    “[Tracking is] like a way of giving myself a reality check because it means that at the end of the day, I am accountable for what I'm doing. So at the end of the day, I can always check to see what I'm doing right or if I'm doing it wrong.” (P23) 
\end{quote}

\textcolor{black}{Given how our participants demonstrated how they self-track their business expenses, income, mileage, and the time they spend working to learn more about themselves, we find that} self-tracking allows gig workers to gather relevant insights about their behavior, activity, and experience. \textcolor{black}{This includes learning what the job could look like for them, what working for the platform is like, and how they are performing in the job}. Having this data increases their self-awareness and autonomy as they can learn from and act on their tracked data.

\subsubsection{Evaluating data to inform work and life}
With enhanced awareness of their performance, gig workers can evaluate their work in alignment with their personal, professional, and financial goals:

\begin{quote}
    “For me, [tracking is] good because it helps you to know if what you're doing is worth it or not, because you don't just work because you want to work. You're working because you want to make a living out of it. [...] So you have to work and have a [record] of what you're doing. If it's not good, if it's not working very well, you can decide to take up a different job.” (P2) 
\end{quote}

Using self-tracking tools allows them to dynamically review and evaluate their data, informing the decisions and assessments they make. Some participants engage in performance assessments by reviewing the efficiency and the viability of their work \textcolor{black}{using data on their income, expenses, mileage, and time. With this data, they can} evaluate “if it is something [they] still need to continue, or if [they are] actually gaining or earning from this kind of work.” (P13). \textcolor{black}{Several participants shared an example highlighting a gig worker's decision-making process, which includes calculating income against expenses like gas and time costs. This becomes crucial when dealing with low-paying, far-distance jobs that may yield a low dollar-per-mile or dollar-per-hour ratio, potentially leading to financial losses due to higher operating costs.} Others tracked and evaluated income from various jobs, especially when working for multiple gig platforms simultaneously (multi-apping): “I also compare with some of my [other jobs] as well. [...] I'm getting \$125 here a day and I'm getting \$110 here.” (P20). From there, participants shared how they reconsidered their involvement with different gig applications:

\begin{quote}
    “Freelance is where I’m spending most of my time because after my analysis, I have found that it is more profitable to me. [...] I’m getting more per hour compared to the Uber jobs. [...] It gives me the rating on my gigs and which gig is better than the other. Just to make my decisions personally.” (P16) 
\end{quote}

With their tracked data, participants can also collaborate with family members to determine if their work is worthwhile. \textcolor{black}{In the case of P10, they reviewed their data with their wife to evaluate} income earned, mileage, and time spent on gig work:

\begin{quote}
    “Sometimes I'll sit down with my wife [...]and take a look at [my spreadsheet] and see. ‘Okay, you work this amount of hours, you made this much. At the end of the day is it worth it?’ [...] She'll take a look at it from a family perspective. ‘Okay, so you made this, you worked this many hours, you traveled this far. Okay, what did you miss? You know our son did this. Our grandkids did this. We could have done this.’ So she'll kind of throw that in there. And that's what I use to evaluate, if that makes sense.” (P10) 
\end{quote}

\textcolor{black}{In summary,} self-tracking fosters a heightened sense of autonomy and control in gig workers’ personal and professional lives as they can make data-driven decisions regarding their jobs using insights from their collected data. This is particularly important as gig work is inherently independent, with no coworkers or managers who can observe their work. Self-tracking provides a sense of organization and control over their jobs, and helps workers maintain accountability for the different responsibilities they may hold.

\subsection{Fiscal and resource accountabilities to the entrepreneurial self}
Gig workers have to consider their obligations and accountabilities as self-employed individuals to both themselves and the government. These include tax preparation, financial management, time management, and vehicle maintenance – all of which encapsulate the fiscal and resource accountabilities of the entrepreneurial selves that gig workers must manage. 

To navigate the multiple accountabilities of self-employed contractors, participants shared how they have to keep track of different aspects of their activity: “I have to be the one doing all the aspects: both the finance, both the planning, both the schedule – kind of everything.” (P19). \textcolor{black}{This includes personally tracking data such as income, expenses, time, and vehicle-related information. While platforms have some data that may be helpful for workers (see Table \ref{tab:tracked-info}), oftentimes gig workers do not find this sufficient for managing their business due to that information being incomplete or being presented ineffectively \textcolor{black}{as shared by our participants and in prior studies} \cite{Cameron_2022, Oei_Ring_2017b, Oei_Ring_2017a}.} The information they track is then used to manage the accountabilities they hold as part of “running a good business” (P9): 

\begin{quote}
    “You know, I'm a company. I don't have an LLC or anything like that, but I'm running a business, so to speak. And part of [that] is keeping track of what's going on. So you know what's going on, you're not guessing.” (P9)
\end{quote}

By tracking different aspects of their daily routines \textcolor{black}{and reviewing summaries or analyses of them (i.e., total profit, dollars per hour)}, gig workers maintain accountability for the various responsibilities related to their status as self-employed contractors.

\subsubsection{Managing and evaluating finances}
Gig work's flexibility, independence, and earning potential make it an appealing option for individuals seeking some source of income \textcolor{black}{\cite{Manyika_etal_2016}}. Self-tracking finances offers workers the ability to fully evaluate the viability of gig work and understand their financial situation: “I just wanted to know if financially I'm doing okay. If the job is paying. So that is why I look at [my data] like, oh am I doing well? Is this business okay?” (P20).

\textcolor{black}{While PI researchers have explored financial tracking \cite{Kaye_etal_2014, Lewis_Perry_2019}, gig workers' approach differs from occupations engaged in personal budgeting and financial tracking due to the precarious nature of gig work \cite{Malin_Chandler_2017} resulting in variable, inconsistent income \cite{De_Stefano_2016}. Given this, gig workers must be diligent in documenting pay, managing deductions, and setting aside money for taxes, as they lack automatic paycheck deductions:}
\textcolor{black}{\begin{quote}
    ”It’s good to be able to calculate the actual amount of money I earned and get out of the trap that these apps put you in. [They] make you appear like you're earning more than you actually are. But having an accurate representation, especially around taxes and wear and tear on my car, tracking that has given me a little bit more reassurance that I'm not just doing this in desperation.” (P15) 
\end{quote}}

\textcolor{black}{While platforms collect and display income information, workers transfer this data to external tracking applications to connect it with their expenses.} Doing so provides them with a comprehensive perspective of their financial standing \textcolor{black}{as they can calculate profits,} helping them manage their money effectively. Participants shared how tracking enables them to review their \textcolor{black}{business} expenses alongside their income \textcolor{black}{to know whether the job is profitable}: 

\begin{quote}
    “It helps me to know what I'm earning and what I'm spending, my expenditure versus my revenue. It makes me very prudent and very reliable with my spending habits. [...] If I'm really getting a profit out of what I'm doing after the whole tax deduction.” (P21)
\end{quote}

\textcolor{black}{To evaluate their profitability on a smaller scale, participants calculated their net income per hour and net income per mile. This helped them visualize their earnings in the face of pay inconsistency: "Some days you only make X amount per hour, and the others you make more. I try to look at it as you would a regular job where you get a week's pay." (P9). Several participants also emphasized the importance of considering fluctuating gas prices, so they calculated net income per mile as a way of managing that instability: "You want to keep track of your profitability [with] dollars per mile. [...] It's only worth your while with gas being as much as it is." (P10).}

\textcolor{black}{Some participants also monitored both work and non-work expenses to assess whether their lifestyles contributed to achieving their profit goals:} “If I feel I have overspent within the month [then] I cut down the expenditure or increase my savings for the [next] month.” (P1). Having this information available to them allows them to proactively adjust their spending behavior and earning activities according to their goals.

\textcolor{black}{In summary, due to the flexibility and independence associated with the gig economy, workers must diligently track their income and expenses to evaluate the profitability and viability of working in the gig economy.}

\subsubsection{Fulfilling tax obligations}
As self-employed contractors, gig workers are independently responsible for calculating, paying, and providing accurate information for their taxes. Participants in this study emphasized the importance of tracking to correctly file their taxes and maximize their income from gig work: “I need to track all of it, so I can do my taxes correctly and get my tax returns done correctly.” (P5). \textcolor{black}{Specifically, they shared how they meticulously tracked gig work-related income, documented business expenses, and recorded business mileage. These practices are crucial to ensuring accurate tax filings and maximizing the potential for tax returns.}

Participants who had worked a W-2 job (a traditional employee role) shared that the necessity of tracking for gig workers can feel exploitative when they know that it isn’t done by everyone:

\begin{quote}
    “It feels like a necessary evil. [...] It's something that I don't have to do at my W-2 job. It's something that people at W-2 jobs don't have to think about.” (P5) 
\end{quote}

In traditional employment roles, employers withhold taxes from employees' paychecks and remit this to the government as a way of fulfilling employees' income tax obligations \textcolor{black}{\cite{Harris_Krueger_2015}}. Meanwhile, independent contractors’ paychecks do not have any income tax withheld, so they have increased responsibility to calculate and pay their income and self-employment taxes \textcolor{black}{\cite{Harris_Krueger_2015}}.

\textcolor{black}{Beyond tracking their activity, gig workers} must also maintain records \textcolor{black}{of information they submit (i.e., income, business expenses, mileage)} in case they are audited by the IRS: “If something did happen where I had to give [the IRS] some documentation, I have this [information] written down right here.” (P9). Participants explained that they are conscious of the potential implications of being audited and, as a result, select a tracking tool that can help them if the event were ever to arise: 

\begin{quote}
    “Shipt [the gig platform] wouldn't track the mileage, so that's why I got Everlance [the tracking app]. [...] With Uber Eats, [...] I'd have so many screenshots and it wouldn't be a perfect breakdown of every order versus Everlance [that] tracks my whole driving session from start to finish. [...] So say, if I got audited and I needed to prove that I drove X amount of miles that I'm claiming on my taxes, if I didn't have Everlance, then I would probably have to go into Uber Eats and go into every order detail and screenshot it.” (P25)
\end{quote}

 \textcolor{black}{Here, P25 shared how they opted to use a third-party tracking tool since the platform they worked on, Shipt, did not track mileage which they knew was needed for their taxes. \textcolor{black}{On the other hand, Uber Eats tracked one-way mileage per job, so it would be extra work on their part to calculate total mileage come tax season.} Other participants shared similar sentiments as they used third-party tracking tools to more accurately track mileage, \textcolor{black}{resonating with prior studies that noted how gig platforms only track mileage when gig workers are actively on the job (i.e., making deliveries or traveling with passengers) \cite{Oei_Ring_2017a, Oei_Ring_2017b}.} Given its importance for taxes, \textcolor{black}{participants emphasized the importance of accurately tracking mileage}: "I will gladly pay money for a tool that guarantees that at the end of the year, I can get that [mileage] deduction accurate." (P7).}

Participants emphasized the pivotal role of tracking for tax purposes in maximizing their profits from gig work, considering it the most important aspect of their job: “If I didn't [track mileage], I would have not made money.” (P7). This is because their tax deductible can “add up” (P5) due to how much mileage they put on for gig work. This emphasis on mileage tracking stems from the significant tax deductions they can claim when reporting mileage and business expenses "otherwise there's a big chunk of money you just give away every year” (P10). The “standard mileage rate” is used to determine deductible costs related to operating a personal vehicle for business purposes \cite{IRS_2022}. Independent contractors who use their personal vehicles for gig work can reduce the taxes they pay with this deductible based on this rate and their total business mileage from the previous year \textcolor{black}{\cite{Oei_Ring_2017b}}.  

Having their tax-relevant information readily accessible and organized on a tracking app is also crucial for workers when it comes to filing their taxes. It not only speeds up the tax process but also enhances the accuracy of the information they enter on their tax forms:

\begin{quote}
    “[Excel] gives me a huge spreadsheet of every drive I made the entire year. [...] I can easily sum up how many miles [I drove] for business versus personal. [...] You get a number and you punch it into your tax software and it spits out a deduction.” (P7) 
\end{quote}

\textcolor{black}{In summary, due to their legal and financial tax obligations, gig workers place high importance on being able to accurately file taxes, maximize tax deductions, and maintain records of their data. To do this, they self-track gig work-related income, business expenses, and business mileage with carefully selected tools.}

\subsubsection{\textcolor{black}{Maintaining vehicles}}
As independent contractors, gig workers supply their own vehicles for the job \textcolor{black}{\cite{Harris_Krueger_2015}}, something especially necessary for rideshare and delivery work. Participants noted the accelerated wear and tear on their vehicles due to the nature of their jobs: “I bought the car in the beginning of August, started ridesharing not long after that. And I've already put on 35,000 miles. Something that people will [do] in a year, I [do] it in just a few months.” (P10). Due to this increased vehicle usage, gig workers also track mileage to guide vehicle maintenance schedules. \textcolor{black}{With this, they prioritize vehicle upkeep and timely maintenance to manage the higher mileage accumulated during gig work}: 

\begin{quote}
    “You want to keep track of how your car is, how it's doing. [...] So, you're gonna be going through the regular wear and tear of the car pretty often. Oil changes, where most people would do it [...] maybe every two to three months or so. On average [for me], every three weeks, more or less.” (P10)
\end{quote}

The increased usage and maintenance costs create a domino effect, leading participants to monitor their car data \textcolor{black}{(e.g., fuel efficiency, mileage, and oil levels)} for more efficient vehicle use. P8 describes tracking and monitoring such data to reduce service center visits and save on maintenance expenses:

\begin{quote}
    “Say on May 1st, [...] so first of the month, I have that much oil life left based on the onboard car computer. It tells me I have 50\%. And then I monitor this situation and check it a month later. So, June 1st, and I saw [it is] 40\%. [...] So, 10\% decrease in a month. So, what did I do? Did I drive harshly or not? I can backtrack and see how I can reduce that number.” (P8)
\end{quote}

\textcolor{black}{In summary,} gig workers heavily rely on their vehicles for their duties, necessitating self-tracking \textcolor{black}{vehicle-related information} for efficient car maintenance scheduling and vehicle utilization. This tracking allows them to maintain accountability and take necessary actions \textcolor{black}{for the upkeep of their vehicles}, enhancing their autonomy as contractors.

\subsubsection{\textcolor{black}{Managing time}}
Gig workers, as independent contractors, must prioritize time management as they are compensated per job, receiving income only upon job completion, unlike traditional employees who receive hourly or annual salaries \textcolor{black}{\cite{Harris_Krueger_2015}}. This underscores the importance of time for gig workers since their earnings depend on the number of completed jobs; efficiency in completing one job enables them to take on the next more quickly:

\begin{quote}
    “Time yourself. Make sure you don’t spend unnecessary time on specific things. Because of the next client that you [have, you] don’t know how many minutes you go pick the person up. [...] Time is money.” (P11)
\end{quote}

\textcolor{black}{Gig workers can track and monitor time through external tools or on the gig platform. However, some gig platforms either do not display working time (Table \ref{tab:tracked-info}) or only track working time. The latter is an issue when workers want to see their total working time: "DoorDash tracks whenever I'm logged on, but TripLog [tracks] from the first thing in the morning to the last thing [in my day]." (P1).} Tracking total time spent working allows workers to evaluate this against their income from a period to understand whether they are spending their time effectively and efficiently as this directly correlates to their income:

\begin{quote}
    “Mostly what I do track is time and [income] so that I know at least during this time to this time I worked on a certain gig and it paid [a] certain amount so that I can get my weekly and monthly accommodations.” (P16) 
\end{quote}

Given the flexibility of gig work, there is also the possibility of \textcolor{black}{being constantly on-call or} working \textcolor{black}{irregular hours}, but this isn’t always the most strategic approach. While gig work offers flexible scheduling, \textcolor{black}{working with a strategy is ideal to prevent aimless work.} Participants shared how they manage their time when engaging with these platforms by setting hourly income goals \textcolor{black}{and identifying the most profitable times to work}:

\begin{quote}
    “The hours spent is more important to me. I really want to maximize the amount of money I make per hour, so I usually work peak[s] – weekends and dinner rushes. [...] I tend to stop working instead of stretching out the time because I find that working more hours isn’t necessarily gonna earn me the rate that I want, it’s picking the right times to do it, that’s generally around the dinner rushes.” (P15)
\end{quote}

\textcolor{black}{Tracking and monitoring how long it takes them to complete a job also helps maintain flexibility in choosing when and how long to work.} This is particularly beneficial for workers who juggle multiple jobs, responsibilities, and commitments. In the case of P19, effective handling of diverse responsibilities is evident through tracking the time they spend on jobs:

\begin{quote}
    “[I track time] when I have other gig work to do and I have a deadline to accomplish it, [but] I still want to do rideshare. So, I dedicate five hours to do the particular work [...] sometimes we get lost in time and we get lost in [...] discussions with clients and stuff like that.” (P19)
\end{quote}

Participants who worked with Amazon Flex also shared how they monitor their time with the platform to optimize earnings: ”I calculate what my running weekly hours worked are because [...] you can't exceed 40 [hours in seven days] otherwise you won't get any more shifts.” (P14). Participants would track their time to make informed choices on which blocks (similar to “shifts” and differ based on pay, length, and number of deliveries) they take: “You want to make sure that you're not taking a low-paying block for three hours. If that puts you at cap, then the next day you can't work a higher-paying three-hour [block].” (P7). 

\textcolor{black}{In summary, the payment structure and flexible work arrangements of gig work commodifies workers’ time. As a result, participants shared how they tracked time to evaluate productivity, maximize dollars per hour, manage multiple jobs, and manage platform limitations.}

\subsection{Performance accountability to the platformized self}
The platformized self refers to the worker’s identity that is influenced by the activity and relationships (i.e., platform, customer) they establish within the gig platform. Workers are lightly managed by the platform through metrics and ratings \textcolor{black}{(Table \ref{tab:platform-metrics})}, which can be either system-generated or customer-sourced \cite{Chan_2022}. \textcolor{black}{Along with metrics, workers may also see the definitions of these metrics (i.e., what they calculate and look at) and infractions they commit; however, it is not always clear when something will count against them and how severely it may affect their metrics.} As algorithmic substitutes for management and performance evaluations, gig workers are held accountable for the contents of their records \cite{Mosseri_2022}, necessitating them to actively understand their metrics, monitor their performance, and maintain a favorable position to avoid consequences or retain benefits. \textcolor{black}{Given this, our participants demonstrated how they self-tracked to understand how platform metrics operate and manage their ratings to avoid consequences.}

\subsubsection{Understanding platform metrics and algorithms}
Metrics play a pivotal role in how gig platforms oversee workers, serving as the primary means of evaluation, with P18 likening these to “an evaluation from your employer.” Poor metrics can be interpreted as “[showing] you might be troublesome, not really dependent, not somebody to trust” (P8). Because of the significant ramifications tied to low metrics \cite{Kirven_2018, Manokha_2020}, participants possessed a clear understanding of their significance and the potential consequences of failing to meet them. Alternatively, having good metrics can show that they are “on a good path” (P19), as workers can be rewarded with advantageous benefits such as better jobs and, therefore, better pay \textcolor{black}{\cite{Chan_2022}}. Nevertheless, participants were attentive to their metrics as they signaled to the platform and customers how they were performing. 

Despite the relevance of platform metrics, information asymmetries exist \textcolor{black}{(i.e., algorithmic opacity and lack of earning/pricing transparency)}, hindering workers from fully understanding how these metrics and algorithms function \textcolor{black}{\cite{Rosenblat_Stark_2016, Jarrahi_Sutherland_2019, Duggan_etal_2019}}. Consequently, workers expressed curiosity regarding the tracking and calculation of their information prompting them to track and calculate whatever information they could \textcolor{black}{(e.g., mileage and income)} to gain insights into platform operations. \textcolor{black}{For instance, \textcolor{black}{resonating with prior work on workers' mistrust of platform-tracked income \cite{Cameron_2022}, P5 shared how they checked whether platforms accurately summed up income data}: "I'll get a calculator and just make sure that they are adding up my income correctly." } \textcolor{black}{Despite this, participants shared how they still faced uncertainty as meanings and algorithms were not always straightforward. While doing simpler calculations allowed them some insight into how platforms calculated straightforward data, other metrics with elements determined by the platform could only be understood through monitoring these metrics’ movements on platforms according to workers’ activities.} For instance, Amazon Flex employs a vague sliding scale system to assess worker performance \textcolor{black}{but is not transparent about what moves the scale up or down}: 

\begin{quote}
    “Those [platform metrics] I mentioned, somehow Amazon lunges it together into a sliding scale of ‘Fantastic’, ‘Great’, ‘Fair’, ‘At Risk.’ [...] I have one delivered and received issue and one on-time delivery issue, and my bar is almost entirely ‘Fantastic’, like I'm just a smidge off the top. But other people will get one delivered and received thing and it will knock them all the way down to ‘Great’. No idea why, no idea how it works. They won't tell us, they never will.” (P7) 
\end{quote}

As workers were unsure how they were rated on this scale, they also faced uncertainty surrounding which behaviors would drop their ratings and potentially strip them of benefits or leave them vulnerable to deactivation: \textcolor{black}{“If it falls down to ‘At Risk’, then the fear is that Amazon will kick you off the platform. Lower than ‘Fantastic’ matters a little bit because Amazon has a rewards program.” (P7)}. Similarly, P10 shared an instance where they noticed their metrics remained unchanged despite accepting more orders, leading to questions about the evaluation process given its implications for their work:

\begin{quote}
    “You go in and you'll do everything you can to get your acceptance rate up, [but] we don't see it going up. I'll go in, and I'll take every single order. Eleven orders and my Acceptance Rate will stay the same. [...] I just question how does [the algorithm] work? I don't understand [it] sometimes.” (P10)
\end{quote}

\textcolor{black}{Acceptance rate is the percentage of received orders that drivers accept \cite{Lee_etal_2015}. However, as P10 did not see this metric changing despite accepting more orders, they began to question how it was really evaluated by platforms.} Relatedly, \textcolor{black}{while reviewing their platform-tracked data,} P3 \textcolor{black}{noticed} declining metrics despite not actively working so took a month off to monitor this. In that period, their metrics continued to drop, \textcolor{black}{causing feelings of distrust: "I believe they manipulated their stats [...] I take that as an extra layer of nonsense that is completely unprofessional." (P3).} As displayed by our participants, the lack of algorithmic transparency impacts the worker-platform relationship by fostering distrust, creating uncertainty, and instilling fear as workers believe companies withhold information without apparent reason. This aligns with prior studies highlighting gig workers' frustration with opaque algorithms \cite{Jarrahi_Sutherland_2019, Rosenblat_Stark_2016}. 

In summary, given the potential consequences of poor metrics, workers displayed heightened attention to them. \textcolor{black}{By self-tracking the same data points as platforms and monitoring their metrics, workers made efforts to understand how algorithms operated and, based on their understanding,} adjusted their behavior to maintain or improve them.

\subsubsection{Acting according to metrics}
Participants actively monitored metrics to safeguard their reputation, adjusting their behavior to maintain good standing and mitigate potential repercussions. Relatedly, Mosseri \cite{Mosseri_2022} discusses “reputation auditing” where gig workers identify and address inaccuracies on gig platforms' records. Sannon et al. \cite{Sannon_etal_2022} also discuss reactive uses of self-tracked data for self-protection, i.e., to contest pay discrepancies, use in pay disputes, and protect against false claims. In addition to this, we found participants proactively monitor their metrics to maintain good standing, even before identifying inaccuracies, and remain vigilant to potential consequences or poor ratings:

\begin{quote}
    “If my rating goes lower, I definitely wanna be careful and try not to do something that I did last time. Just to be sure that it stays there. Same with the acceptance rate because if you decline too many orders, if your acceptance rate goes below 50\%, then you're not prioritized for high-paying orders. So, it definitely keeps you in check. At least for me personally, I look at it all the time.” (P4)
\end{quote}

\textcolor{black}{To proactively monitor metrics,} another participant described taking notes of their deliveries to remember instances that might negatively impact their metrics. These notes \textcolor{black}{specified incidents that prevented the worker from completing a delivery and what consequence they anticipated, for example}:

\begin{quote}
    “I had one [delivery] where a business was closed [...] because the entire office got COVID. So I made a note just so I can remember to check my standings in a few days to see if it counted against me.” (P7)
\end{quote}

\textcolor{black}{These served as bookmarks for potential issues such as not receiving payment or getting deliveries marked late or missing. As P7 put it: “things that are abnormal that Amazon might come and yell at me about.” In these instances, the participant was aware of actions that may result in infractions in the eyes of the platform, potentially impacting their metrics.} By being proactive about preventing poor metrics, participants demonstrated how they avoid falling into a bad position which is crucial as this has implications on their income and experience with gig work. 

Participants also shared how they use tracked data \textcolor{black}{(e.g., time they completed a job)} and photos \textcolor{black}{of successfully delivered packages} during disputes, similar to prior work on reputation auditing \cite{Mosseri_2022}. In addition, we found that if workers can’t successfully dispute a mark against them, they may opt to wait for its removal from their record. Certain platforms like Amazon Flex automatically remove marks after a certain number of deliveries, prompting workers to track and monitor their deliveries to “gauge of how long it’s going to take to work those issues off [their] standings” (P7).

In summary, participants displayed a heightened awareness of their metrics and the associated consequences. As those being evaluated, they are then accountable for maintaining favorable metrics, whether through continuous monitoring for maintenance or disputing with self-tracked data. This behavior stems from concerns about income loss from platform-sanctioned penalties (e.g., receiving unfavorable orders) or job insecurity (e.g., deactivation). Leveraging self-tracking enabled them to proactively manage metrics by increasing awareness and holding themselves accountable for their performance, customer satisfaction, and platform compliance.
\section{Discussion}
In our findings, we illustrated how gig workers self-track to maintain accountability across three distinct identities. First is the holistic self, where they maintain personal accountabilities of learning their capabilities in gig work, managing their work/life balance, and acting in their self-interest. Second is the entrepreneurial self, characterized by their being self-employed individuals managing fiscal and resource accountabilities. Last is the platformized self, as they navigate and negotiate performance-related accountabilities with the platform and customers. \textcolor{black}{Participants shared how due to insufficient platform tracking (summarized in Table \ref{tab:insuff-data}) they self-tracked various types of data (summarized in Table \ref{tab:tracked-data}) to manage these different accountabilities.} Below, we reflect on the accountable self that gig workers embody and the implications of our findings as they relate to the invisible labor of self-tracking and asymmetries present in gig work. We then offer design implications to support self-tracking for gig workers’ multi-dimensional accountability.

\aptLtoX[graphic=no,type=html]{
\begin{table*}[b]
 \centering
 \caption{Summary of insufficient data tracking or presentation by gig working platforms.}~\label{tab:insuff-data}
 
  \begin{tabular}{|p{5cm}| p{5cm} | p{5cm}|} 
 
\hline
\rowcolor[HTML]{e6e6e6}
{ \textbf{Expenses}}
 & { \textbf{Income}}
 & { \textbf{Mileage}}
 \\ 
 \hline

 
Personal and business expenses are not tracked by the gig platform. 

& – Income is often separated by day or task, requiring extra effort on the worker's side to transfer this information and calculate or view it in different periods (i.e., weekly, monthly, per hour). \hfill\break

– Income may be miscalculated by the platforms.
& – If tracked by the platform (Table 2), mileage is often separated by day or job, requiring extra effort on the worker's side to transfer this information to be able to calculate or view it in different periods (i.e., weekly, monthly, per hour).\hfill\break

– Only calculated when the driver is actively working on an assignment; not calculated by the platform when drivers are driving and waiting to get a new assignment. 
\\ 
 \hline

\hline
\rowcolor[HTML]{e6e6e6}
 { \textbf{Working Time}}
 & { \textbf{Qualitative Experiences}}
 & { \textbf{Platform Metrics}}
 \\ 
 \hline
 
If tracked by the platform (Table 2), time spent working is often separated by day or job, requiring extra effort on the worker's side to transfer this information to be able to evaluate it alongside other factors (i.e., mileage, income). 
& Qualitative experiences of the worker (e.g., personal accounts and notes on deliveries or clients) are not collected by the gig platform. 
& There is a lack of transparency in how metrics are tracked, calculated, and used by the platform. \\ \hline

 \end{tabular}
\end{table*} }{
\begin{table*}[b]
 \sffamily\smaller
 \def\arraystretch{1.3}
 \centering
 \caption{Summary of insufficient data tracking or presentation by gig working platforms.}~\label{tab:insuff-data}
 
\scalebox{0.88}{
  \begin{tabular}{!{\color{black}\vrule} L{0.34\textwidth} | L{0.34\textwidth} | L{0.34\textwidth}!{\color{black}\vrule}} 
 
\arrayrulecolor{black}\hline
\rowcolor{lightGrey}
{\smaller \textbf{Expenses}}
 & {\smaller \textbf{Income}}
 & {\smaller \textbf{Mileage}}
 \\ 
 \hline

 
Personal and business expenses are not tracked by the gig platform. 

& – Income is often separated by day or task, requiring extra effort on the worker's side to transfer this information and calculate or view it in different periods (i.e., weekly, monthly, per hour). 

– Income may be miscalculated by the platforms.
& – If tracked by the platform (Table 2), mileage is often separated by day or job, requiring extra effort on the worker's side to transfer this information to be able to calculate or view it in different periods (i.e., weekly, monthly, per hour).

– Only calculated when the driver is actively working on an assignment; not calculated by the platform when drivers are driving and waiting to get a new assignment. 
\\ 
 \hline

\arrayrulecolor{black}\hline
\rowcolor{lightGrey}
 {\smaller \textbf{Working Time}}
 & {\smaller \textbf{Qualitative Experiences}}
 & {\smaller \textbf{Platform Metrics}}
 \\ 
 \hline
 
If tracked by the platform (Table 2), time spent working is often separated by day or job, requiring extra effort on the worker's side to transfer this information to be able to evaluate it alongside other factors (i.e., mileage, income). 
& Qualitative experiences of the worker (e.g., personal accounts and notes on deliveries or clients) are not collected by the gig platform. 
& There is a lack of transparency in how metrics are tracked, calculated, and used by the platform. \\ \hline

\arrayrulecolor{black}\hline
 \end{tabular}
}
\end{table*}}
\aptLtoX[graphic=no,type=html]{
\begin{table*}[htbp]
 \centering
 \caption{Summary of data and analyses gig workers use to manage their different accountabilities.}~\label{tab:tracked-data}
 \begin{tabular}{ |p{3cm} | p{4.5cm}|  p{3cm}| p{5cm} | p{3cm} | p{4cm}|}
\hline
\multicolumn{2}{|c|}{\cellcolor[HTML]{e6e6e6}{\textbf{Personal Accountabilities}}} & \multicolumn{2}{c|}{\cellcolor[HTML]{e6e6e6}{\textbf{Financial and Resource Accountabilities}}} & \multicolumn{2}{c|}{\cellcolor[HTML]{e6e6e6}{\textbf{Performance Accountabilities}}}
 \\ 
 \hline

\rowcolor[HTML]{e6e6e6}{ \textbf{Purpose}} & { \textbf{How}} & { \textbf{Purpose}} & { \textbf{How}} & { \textbf{Purpose}} & { \textbf{How}}
 \\ 
 \hline
To learn what the job and platform are like & – Calculating profit (income, expenses), dollars per mile (income, mileage), and dollars per hour (income, time) to know their earning capacity\hfill\break

– Tracking expenses, income, mileage, and time to identify patterns in data 
& To determine the profitability and sustainability of their work & – Tracking income and expenses to calculate how much they are earning (income, expenses), after considering financial expenses. \hfill\break

– Calculating dollar per hour (income, time) and dollar per mile (income, mileage) to evaluate profitability on a smaller scale
& To understand how platform algorithms calculate certain information & Personally tracking mileage and income to see if these are being fairly calculated by the platform \\ \hline
To know if a job is worth it & – Identifying profit (income, expenses) to see if they are earning money from the job\hfill\break

– Calculating dollars per hour (income, time) or dollars per mile (income, mileage) to know if they are earning enough per hour to sustain their financial needs 
& For tax needs (i.e., filing, calculating deductions, audits) & – Tracking and keeping records of income, business expenses, and mileage, as required by the IRS\hfill\break

– Tracking and keeping records of business expenses and mileage, as these can be used for tax deductions 
& To dispute poor metrics (i.e., falling below a set threshold) which can lead to consequences (i.e., deactivation) & Tracking personal experiences while working with photos and notes (i.e., on businesses and restaurants) to provide additional documentation for job completion or to explain reasons for non-completion \\ \hline

To manage work/life balance & – Tracking time to set boundaries on work\hfill\break

– Calculating profit (income, expenses) to track their progress on financial goals 
& To monitor vehicle wear and tear & Tracking vehicle maintenance schedules, mileage, oil levels, fuel efficiency, and other vehicle-related information to stay on top of routine maintenance tasks and ensure optimal vehicle performance 
& To estimate when infractions (e.g., late or missing deliveries) will be removed from their records & Tracking the total number of orders within a certain period to recognize when they have improved their standing with platforms \\ \hline

{} & {}
& To manage time between other jobs or personal commitments & Tracking and monitoring time to identify how long they have been working particular jobs or to set limits on working 
& To monitor and review performance to prevent poor metrics (i.e., falling below a set threshold) & Monitoring platform metrics to be able to adjust behavior before metrics worsen \\ \hline
{} & {} 
& To monitor productivity and efficiency & – Calculating dollars per hour (income, time) to assess the value of their time and resources in relation to their overall output and income\hfill\break

– Tracking how much time they spend working and what they get done to ensure they are using their time wisely & {} & {} \\ \hline

\hline
 \end{tabular}
\end{table*} }{
\begin{table*}[htbp]
 \sffamily\smaller
 \def\arraystretch{1.3}
 \centering
 \caption{Summary of data and analyses gig workers use to manage their different accountabilities.}~\label{tab:tracked-data}
\scalebox{0.88}{
 \begin{tabular}{!{\color{black}\vrule}  
 L{0.14\textwidth} | L{0.18\textwidth} | L{0.14\textwidth} | L{0.18\textwidth} | L{0.14\textwidth} | L{0.18\textwidth}
 !{\color{black}\vrule}} 
 
\arrayrulecolor{black}\hline
\rowcolor{lightGrey}

\multicolumn{2}{|c|}{\textbf{Personal Accountabilities}} & \multicolumn{2}{c|}{\textbf{Financial and Resource Accountabilities}} & \multicolumn{2}{c|}{\textbf{Performance Accountabilities}}
 \\ 
 \hline
 
 \arrayrulecolor{black}\hline
\rowcolor{lightGrey}

 {\smaller \textbf{Purpose}} & {\smaller \textbf{How}} & {\smaller \textbf{Purpose}} & {\smaller \textbf{How}} & {\smaller \textbf{Purpose}} & {\smaller \textbf{How}}
 \\ 
 \hline

To learn what the job and platform are like & – Calculating profit (income, expenses), dollars per mile (income, mileage), and dollars per hour (income, time) to know their earning capacity

– Tracking expenses, income, mileage, and time to identify patterns in data 
& To determine the profitability and sustainability of their work & – Tracking income and expenses to calculate how much they are earning (income, expenses), after considering financial expenses. 

– Calculating dollar per hour (income, time) and dollar per mile (income, mileage) to evaluate profitability on a smaller scale
& To understand how platform algorithms calculate certain information & Personally tracking mileage and income to see if these are being fairly calculated by the platform \\ \hline

To know if a job is worth it & – Identifying profit (income, expenses) to see if they are earning money from the job

– Calculating dollars per hour (income, time) or dollars per mile (income, mileage) to know if they are earning enough per hour to sustain their financial needs 
& For tax needs (i.e., filing, calculating deductions, audits) & – Tracking and keeping records of income, business expenses, and mileage, as required by the IRS

– Tracking and keeping records of business expenses and mileage, as these can be used for tax deductions 
& To dispute poor metrics (i.e., falling below a set threshold) which can lead to consequences (i.e., deactivation) & Tracking personal experiences while working with photos and notes (i.e., on businesses and restaurants) to provide additional documentation for job completion or to explain reasons for non-completion \\ \hline

To manage work/life balance & – Tracking time to set boundaries on work

– Calculating profit (income, expenses) to track their progress on financial goals 
& To monitor vehicle wear and tear & Tracking vehicle maintenance schedules, mileage, oil levels, fuel efficiency, and other vehicle-related information to stay on top of routine maintenance tasks and ensure optimal vehicle performance 
& To estimate when infractions (e.g., late or missing deliveries) will be removed from their records & Tracking the total number of orders within a certain period to recognize when they have improved their standing with platforms \\ \hline

{} & {}
& To manage time between other jobs or personal commitments & Tracking and monitoring time to identify how long they have been working particular jobs or to set limits on working 
& To monitor and review performance to prevent poor metrics (i.e., falling below a set threshold) & Monitoring platform metrics to be able to adjust behavior before metrics worsen \\ \hline

{} & {} 
& To monitor productivity and efficiency & – Calculating dollars per hour (income, time) to assess the value of their time and resources in relation to their overall output and income

– Tracking how much time they spend working and what they get done to ensure they are using their time wisely & {} & {} \\ \hline

\arrayrulecolor{black}\hline
 \end{tabular}
}
\end{table*}}

\subsection{The accountable self in gig work}
Our findings describe how gig workers embody an ‘accountable self’ as they have “the capacity and willingness […] to give explanations for [their] conduct” \cite{Masiero_2020}. \textcolor{black}{Their faceted identities are organized considering gig workers' simultaneous  involvement in multiple roles with distinct responsibilities \cite{Caza_Wilson_2009}. Faceted identities are influenced by social context as individuals play diverse roles, navigate various relationships, and manage distinct expectations in different social settings \cite{Farnham_Churchill_2011, DiMicco_Millen_2007}. In the context of gig workers' holistic and entrepreneurial selves, these identities emerged due to the unique responsibilities they bear toward themselves and external stakeholders (e.g., family, government). Furthermore, participants in our study referred to the gig work they do as a business, seeing it as both an activity and a distinct business entity, aligning with neoliberalism, a socio-economic ideology that views the self as an enterprise that is “both a member of the firm and as itself a firm” \cite{Brown_2015}. Within this framework, individuals work within organizations and operate as independent enterprises, actively managing various business-related accountabilities \cite{McNay_2009}, i.e., fiscal and resource accountabilities of the entrepreneurial self. Meanwhile, gig workers' platformized selves can be seen as a form of performance of self-presentation \cite{Goffman_1959}, as they must navigate outward-facing metrics visible to algorithms and customers \cite{Munoz_etal_2022}.} 

\begin{figure*}
    \centering
    \includegraphics[width=0.65\textwidth]{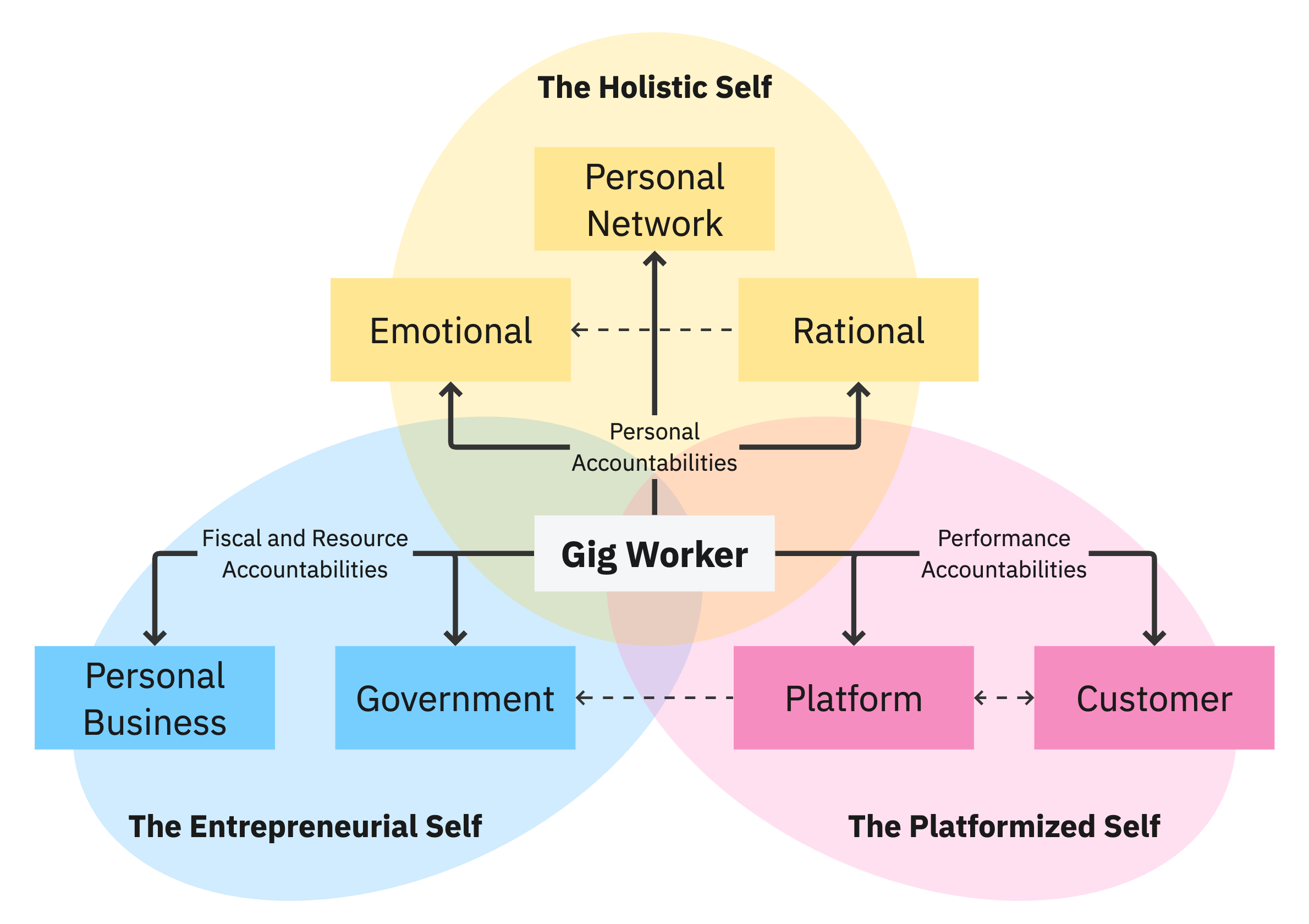} 
    \caption{\small{The Accountable Self of Gig Workers. Continuous lines represent gig workers' accountability to various parties, while broken lines depict potential accountability among these parties. Within the holistic self, the rational self is accountable for the emotional self, as discussed in 5.1. Platforms are accountable to the government for legal and regulatory compliance. Platforms and customers hold mutual accountability, primarily focused on trust and reliability.}}
    \label{fig:accountable-self}
    \Description{This figure illustrates three key aspects of gig workers, based on our research: the holistic self, the entrepreneurial self, and the platformized self. Each aspect is represented by a circle in a Venn diagram, with all three circles overlapping to represent the gig worker as a whole.Starting from the center, we show how gig workers are accountable to different aspects of themselves. First, they have personal accountabilities to their holistic self, which includes their personal network, emotions, and rationality. We use broken lines to indicate that their rational self is accountable for their emotional self.Next, gig workers have financial and resource accountabilities to their entrepreneurial self, which includes accountabilities to their personal business and to the government. Lastly, they have performance accountabilities to the platformized self, which includes accountabilities to the customer and the platform. We use broken lines to represent mutual accountabilities between the platform and customers, as well as the platform's accountability to the government.Overall, this figure highlights the multifaceted nature of gig work and the various accountabilities that gig workers must balance.}
\end{figure*}

Understanding how gig workers navigate their accountabilities within a network of stakeholders (Figure \ref{fig:accountable-self}) highlights the dynamics within the gig economy. This knowledge is relevant for academic exploration and has practical implications for policymakers, businesses, and labor advocates seeking to address the challenges and opportunities presented by this landscape. 

Within the accountable self, gig workers display accountability to both external entities and themselves. In being self-accountable, individuals assume responsibility for their actions and decisions without external observation or enforcement \cite{Ghanem_2022}. The \textcolor{black}{independence and atomization of gig work \cite{Yao_etal_2021}} led to primarily self-driven \textcolor{black}{and self-managed} accountabilities across their various identities. \textcolor{black}{While prior literature has focused more generally on self-management, i.e., studying how self-management is applied across various activities and goals \cite{Alvarez_etal_2022, Alvarez_etal_2023, Hsieh_etal_2022, Dose_Klimoski_1995, Ghanem_Castelli_2019}, we focused on a specific form of self-management where gig workers use self-tracking to maintain accountability.} 

In the holistic self, gig workers demonstrate self-accountability as their rational selves are accountable for maintaining their emotional selves by maintaining an awareness of their activity, evaluating and reflecting on their performance, acknowledging their achievements, and considering various perspectives on their work. In the entrepreneurial self, gig workers, as self-employed individuals, are self-accountable to their businesses as a separate entity from themselves. However, even when gig workers are accountable to external parties, explanations or evaluations of one’s work are not imminent, underscoring the role of self-accountability in managing gig workers’ obligations. For example, regarding tax responsibilities, repercussions typically surface in the event of an audit. This necessitates gig workers’ self-accountability in proper and timely tax filing. Similarly for the platformized self, disputing poor metrics for performance accountability relies on workers initiating the dispute process. Resonating with Ghanem \& Castelli's \cite{Ghanem_Castelli_2019} findings on the connection between identity and self-accountability in business leaders, we find that gig workers’ external accountabilities (i.e., to the government, platforms, and customers) increase their felt self-accountability as they recognize the importance of meeting these external obligations to maintain their income and job. 

Our findings underscore the multi-dimensional nature of gig workers’ tracking-mediated accountabilities. Participants used self-tracking to manage the multi-dimensional accountabilities they bear. This differs from the focus of existing PI research which has focused on tracking or affecting a singular aspect of the self, i.e., health \cite{Figueiredo_etal_2018}, fitness \cite{Tong_etal_2016}, or productivity \cite{Kim_etal_2019}. While researchers have also explored integrating multiple sources of data, these similarly focus on deriving insights that are relevant to a singular dimension, i.e., health \cite{Lin_etal_2012} or wellbeing \cite{Bentley_etal_2013}. PI researchers have also considered how self-tracking systems can hold individuals accountable for their behavior change goals \cite{Epstein2015Lived} through the social accountability that manifests through publicly posting one’s own tracked data \cite{Munson_Consolvo_2012, Newman_etal_2011, Wang_etal_2022}, the support they receive from posting \cite{Consolvo_etal_2006, Epstein_etal_2016, Munson_Consolvo_2012}, and in-app rankings \cite{Gui_etal_2017}. In contrast, our findings illustrate that gig workers employ self-tracking to simultaneously manage a broad spectrum of accountabilities that extend beyond holding oneself responsible for behavioral change goals. Our participants felt accountable for a varied group of people including themselves, their families, and other entities. Our findings also expand on accountability research that largely emphasizes individuals’ obligations to external parties, which is the more traditional view of accountability \cite{Ghanem_Castelli_2019}. For gig workers, self-tracking activities are not solely about external obligations; they also underscore accountability to oneself.

Although gig workers must manage a wide spectrum of accountabilities, our findings indicate that gig platform designs inadequately support gig workers’ accountability identities. \textcolor{black}{Participants consistently highlighted instances where gig platform tracking falls short in capturing information crucial for maintaining accountability across various facets of their work (Table \ref{tab:insuff-data}). As a result, participants sought external tracking tools to fill these gaps. With these tools, participants could track and sum up data from the entire period they work and test how apps track or even manually input data to ensure the accuracy and comprehensiveness of tracked data.} Having to self-track thus places a burden on workers to invest time in selecting and customizing tools to meet their accountability management needs, \textcolor{black}{such as in the case of P25.} This can be seen as a transfer of accountabilities primarily driven by the absence of adequate design support for the multi-dimensional tracking needs of gig workers. Considering gig workers must navigate multiple dimensions of accountability, external tools become necessary to address this complex accountability landscape. We expand on designing for multi-dimensional accountability management on self-tracking tools in 5.4.

\subsection{Self-tracking as invisible labor}
Our research highlights gig workers' self-tracking practices as a form of invisible labor\textcolor{black}{, emphasizing the significant effort and investment inherent in this activity. As our findings indicate, gig workers have to expend considerable effort in tracking due to their multi-dimensional accountabilities and the insufficiency of gig platforms in supporting these needs with adequate data, as summarized in Table \ref{tab:insuff-data}. To manage their accountabilities, gig workers must carefully select tools that fit their specific needs. They are also responsible for identifying the specific data to track and conducting meticulous analyses, both of which are summarized in Table \ref{tab:tracked-data}. Additionally, they manage other tracking-related responsibilities such as retaining records and reviewing past performance. Despite the considerable effort invested in self-tracking practices, gig workers remain uncompensated for the additional workload they entail.} 

Gig workers are primarily paid on a per-task basis, which means they are not compensated for managing activities beyond the gig task. This payment structure perpetuates issues of unpaid, invisible labor necessary to participate in the gig economy, e.g., emotional and temporal labor \cite{Raval_Dourish_2016}, self-protective strategies \cite{Sannon_etal_2022}, and work engagements \cite{Toxtli_etal_2021}. As independent contractors, gig workers see accountabilities that are usually handled by one’s place of employment transferred to the individual. These include ensuring tax compliance, covering business expenses, maintaining personal vehicles for work, and managing work schedules, all of which \textcolor{black}{can be self-managed with self-tracking. While previous studies on independent contractor self-management have largely focused on strategizing \cite{Alvarez_etal_2022, Alvarez_etal_2023, Hsieh_etal_2022}, we also found gig drivers self-manage by tracking to fulfill entrepreneurial self and platformized self accountabilities. Here, the concept of self-tracking shifts from a managerial standpoint to viewing the worker at the intersection of labor, datafication, and self-actualization. }

 As gig workers self-track to manage shifted accountabilities, we observed how self-tracking is done to navigate the unclear boundary between contractors and employees as platforms do not provide sufficient tracking features to support the management of multi-dimensional accountabilities. In the U.S., the murky boundary between employee and contractor that gig workers navigate is rooted in the more commonly used binary methods of classifying workers as either independent contractors or employees \cite{Rauch_2021}. \textcolor{black}{This poses a problem as gig workers share similarities with both employees and independent contractors \cite{Harris_Krueger_2015}}. Gig platforms benefit from classifying gig workers as contractors as they don’t have to provide mandated benefits and protections such as health insurance, minimum wage, and unemployment benefits \cite{Cherry_2023, Rauch_2021}. One of the key factors in the most commonly used worker classification tests in the U.S. (Common Law Test, ABC test, Economic Realities test) for determining worker classification is the degree of control and independence a worker has \cite{NFIB_2020, Provenzano_2023, Rauch_2021}. Some argue that platform metrics essentially control workers similar to traditional employees, evident in pervasive algorithmic management and the fostered economic dependency on gig work \cite{Clark_2020}. Participants in our study demonstrated how they manage these forms of control by understanding and managing their performance metrics (platformized self) and tracking finances to maximize their income (entrepreneurial self). While outdated, binary methods of classifying workers that don't effectively align with the gig economy structure \cite{Malik_2017, Provenzano_2023, Woodcock_Graham_2020, Harris_Krueger_2015} persist, it's likely that we will see minimal changes in how gig workers are compensated. However, a transformation in these systems is unfolding as several cases have been raised surrounding the classification of gig workers with some successfully reclassifying workers as employees and others granting them employee-like benefits \cite{Cherry_2023, Woodcock_Graham_2020, Harris_Krueger_2015}. 
 
 Reclassifying workers to receive benefits similar to employees could transform how they self-track. \textcolor{black}{If gig workers were to operate more like employees rather than business owners, tracking for their entrepreneurial selves could change as they receive a more stable income, more consistent hours, and would potentially use company-owned resources for work. Harris \& Krueger \cite{Harris_Krueger_2015} propose an independent worker classification as a "middle ground between traditional employment and independent contractors," offering increased protections and benefits comparable to those enjoyed by employees, which could have implications on gig workers' self-tracking. First, platform companies withholding taxes could distribute tax accountability, reducing the administrative burden of tax filing. This could bring changes to the timing of tax payments, modifications to income calculation, and a less stressful tax management process that currently relies on workers' self-tracking. Second, the implementation of wage and hour protections could potentially reshape how gig workers track and manage their income and time, introducing more structure and predictability. In summary, reclassifying gig workers could reduce their accountabilities, lessening the necessity for extensive tracking.}

 Our findings also depict self-tracking as a burdensome necessity for individuals to protect and preserve their jobs, underscored by the legal and financial implications of not tracking. We found that gig workers track their income and mileage to stay accountable for their financial and tax-related obligations. P5 highlighted the burden of tracking, a task absent in traditional workplace employment but necessary in gig work. Failure to track jeopardizes tax filing accuracy and income optimization, resulting in financial losses and the risk of IRS audits and fines. By putting individuals in charge of their businesses, the gig economy seems to encapsulate the ideal of neoliberalism, which centers on government deregulation and emphasizes personal responsibility \cite{Gerstle_2022}. However, it fails to acknowledge the power imbalances inherent in the gig economy notably skewed in favor of the platform. Workers thus face disadvantages as they are not fairly compensated for essential work that sustains their personal business and the platform ecosystem. The gig economy, thus, challenges the sustainability of neoliberalism as workers see increased accountabilities which they are not fairly compensated for.

  Given the issue of unnoticed and uncompensated self-tracking labor, we would like to highlight worker advocacy groups like the National Domestic Workers Alliance Gig Worker Advocates~\footnote{\url{https://www.gigworkeradvocates.org/}} which prioritize fair compensation for gig workers. Advocacy in the gig economy can unveil the invisibility of these practices, pushing for recognition and compensation of essential tasks beyond gigs to sustain their jobs and adhere to their legal duties.

\subsection{Self-tracking to mitigate information and power asymmetries}
\textcolor{black}{Self-tracking holds a nuanced role for gig workers. While highlighting the issue of uncompensated labor and the broader structural issues within gig work platforms is crucial, it's worth noting that tracking can be a means of empowerment and negotiation for workers as they mitigate the information and power asymmetries prevalent in the gig economy.} Platforms notably withhold useful information (i.e., algorithmic payment determinations, job assignments, performance evaluations) to maintain control over workers \cite{Duggan_etal_2019, Jarrahi_Sutherland_2019, Maffie_2023, Möhlmann_Zalmanson_2017, Rosenblat_Stark_2016}. We found that as gig workers track their performance and activity, they generate knowledge about themselves and the platforms they work for, helping them mitigate pervasive information and power asymmetries. 

Tracking offers workers an objective overview of their performance and activity, enabling self-assessment. This is beneficial as platforms have deemphasized the true costs of platform work and emphasized gross earnings without considering gas, maintenance, and depreciation \cite{Maffie_2023}. An accurate and historical view of their performance \textcolor{black}{(i.e., reviewing income in relation to time spent, miles driven, or jobs done)} lets workers know what they are capable of doing and earning with gig work. With access to historical earnings data, they can gain insights into a platform's profitability, equipping them to decide whether to work more or less with that platform. For instance, P14's consistent income tracking revealed pay rate fluctuations and empowered them to make an informed decision about their engagement with that platform. The knowledge workers gain can mitigate information asymmetries related to pay and earning transparency as it provides insights into performance trends, allowing them to plan their work hours and make job choices based on past performance data. This grants them both an objective performance perspective and insights into the gig platforms, empowering them to make informed decisions about their work lives.

 As workers gain knowledge of themselves and their jobs \textcolor{black}{through self-tracking (Table \ref{tab:tracked-data})}, they gain autonomy and control over their professional lives. While platform metrics control and hold workers accountable, external self-tracking allows workers to subvert this control to some extent, e.g., disputing customer and platform issues for self-protection \cite{Mosseri_2022, Sannon_etal_2022} \textcolor{black}{or collectively demystifying blackbox algorithms \cite{Calacci_Pentland_2022, Stein_etal_2023}}. Thus, self-tracking is useful but necessary as a form of self-preservation and self-protection. This is despite the observed autonomy paradox in gig work that contrasts the expectations of autonomous and flexible work arrangements against the reality of workers experiencing surveillance, control, and a lack of access to information \cite{Jarrahi_etal_2020, Maffie_2023, Möhlmann_Zalmanson_2017}. The enhanced autonomy they experience draws similarities to PI studies where individuals are empowered to manage their health using self-tracking \cite{Figueiredo_etal_2018, Potapov_Marshall_2020}. 

However, it's crucial to note that workers cannot completely eliminate information and power asymmetries through self-tracking; rather, tracking empowers them to address and alleviate challenges only to some extent. Power imbalances persist not only within gig work structures but also extend into the design of platform metrics, stemming from broader contexts such as legal and economic frameworks. Our findings highlight the insufficient support provided by platforms \textcolor{black}{(Table \ref{tab:insuff-data})}, underscoring existing power imbalances stemming from the lack of worker-centered design on gig platforms’ metrics and tracked data. Platform metrics replace human supervisors as performance evaluators and, thus, primarily exist to surveil workers rather than enable them to make informed decisions, underscoring the autonomy paradox in gig work. Gig platforms prioritize control and compliance – showcasing surveillance-centered design – over supporting workers' autonomy and wellbeing, i.e., constant monitoring \cite{Alvarez_etal_2023}, punitive practices \cite{Chan_2022}, and a lack of transparency \cite{Jarrahi_Sutherland_2019}. 

This underscores the significance of worker rights advocacies like Gig Workers Rising~\footnote{\url{https://gigworkersrising.org/}}  which strive to establish enhanced legal and economic structures including protections and benefits like health insurance, job security, wage protection, and fair and transparent labor practices and policies. \textcolor{black}{These causes relate to projects like Shipt Calculator \cite{Calacci_Pentland_2022} that seek to empower gig workers through collective data tracking and sharing to increase wage transparency. The tool employs collective self-tracking to audit and deconstruct the algorithmic blackbox, which could shed light on additional ways workers can empower themselves and mitigate existing asymmetries through tracking.} \textcolor{black}{Current individualized self-tracking tools also complement efforts for data collectives as a way to address asymmetries in the gig economy \cite{Micheli_etal_2020, Stein_etal_2023, Ho_Chuangt_2019, Zhang_etal_2022}. While individual self-tracking empowers workers to manage their affairs and gain personal insights, data collectives offer the opportunity to view a broader perspective. This can reveal industry-wide trends, highlight common challenges faced by gig workers, and serve as a powerful tool for advocacy and collective bargaining. By considering how self-tracking can support advocacy work or data collective efforts, there is potential to utilize existing activities to not only improve the conditions of individual gig workers but also contribute to systemic changes within the gig economy, fostering a more equitable and transparent future.}

\subsection{Design implications for supporting gig workers' multi-dimensional tracking}
Our findings detail how \textcolor{black}{it can be burdensome for} gig workers to self-track to independently manage multi-dimensional accountabilities. \textcolor{black}{This is exemplified by the necessity of tracking to meet tax and financial obligations, and the task of selecting and customizing a variety of tools and data for tracking and analysis.} However, existing PI literature predominantly focuses on tracking singular dimensions of the self. In light of this, we offer design implications for supporting multi-dimensional self-tracking. We propose worker-centric self-tracking systems to empower gig workers in managing various accountabilities, prioritizing worker needs and autonomy in design. Prior research on worker-centered design in the gig economy discussed incorporating workers’ visions into platform design \cite{Qadri_D’Ignazio_2022}, addressing the multifaceted challenges gig workers face \cite{Ma_etal_2023}, and democratizing algorithmic design \cite{Calacci_Pentland_2022}. We discuss design implications related to data collection, analysis, and transparency in both gig platform-based and third-party tracking systems \textcolor{black}{to reduce the burden of multi-dimensional accountability management.}

Design efforts are needed to explore customizable data collection methods, allowing gig workers to adapt them to their unique goals, needs, and interests. This encompasses choosing \textcolor{black}{what to track (i.e., income, job details), how to track (i.e., degree of detail), data sources (i.e., gig platform, third-party app), and acquisition methods (i.e., automatic, manual). This customization is particularly relevant to the three dimensions of the self that gig workers are accountable to as individuals can tailor their tracking based on their priorities. For example, someone more curious about understanding platform dynamics might collect more data related to their performance accountabilities.} Additionally, participants highlighted the use of multiple tracking tools for various information. Importing and integrating data from diverse sources can enhance the generation of comprehensive insights, support informed decision-making, and address gig workers' diverse tracking needs, \textcolor{black}{thereby reducing input and analysis efforts. The implementation of comprehensive integration features can streamline the process, reducing the time and effort required for gig workers to select and tailor their tools and data.}

Designing for flexible data analysis could also empower individuals to extract relevant insights from their data based on their preferences. Participants collected diverse data types including quantitative measurements (i.e., mileage, time, income), qualitative notes (i.e., notes on deliveries or clients), and multimedia content (i.e., map data, photos). Analyzing multiple data types alongside one another can enhance their ability to interpret their data within its broader context\textcolor{black}{, easing the effort associated with combining different data from various sources}. Furthermore, the customizability of data analysis would enable workers to select what data they analyze together and how they will be analyzed. Various data visualization options (i.e., different time frames and visualization types) could accommodate users' unique needs, interests, and data literacy levels, facilitating a deeper understanding of their data and the identification of meaningful patterns. \textcolor{black}{The inclusion of more flexible data analysis features could also enhance support for gig workers as they navigate the different aspects of their worker identity, as detailed in this study. By combining relevant data and subsequently highlighting elements most crucial to their needs and interests, this approach caters to their diverse priorities. Whether they are wellbeing-related considerations for their holistic self, business-related insights for their entrepreneurial self, or metric-related observations for their platformized self, this recommendation addresses the multifaceted dimensions of their work. Overall, the incorporation of customizable data analysis methods not only reduces the time-consuming aspects of self-tracking but also facilitates a more nuanced understanding of their data, enabling gig workers to identify meaningful patterns with greater ease.}

Finally, offering data transparency could support individuals' understanding of how their information is handled and used. For platform metrics\textcolor{black}{, and of relevance to their platformized self,} this entails understanding how and why their metrics are tracked. Our study participants expressed uncertainty regarding the purpose and calculation of platform metrics, which affected workers' ability to evaluate their performance according to the platforms' standards. Furthermore, platforms and third-party tools should also consider transparency surrounding how they handle (i.e., collect, store, use) and process (i.e., protect, anonymize) data. The knowledge this transparency would provide can increase worker agency as they understand the implications of being monitored and empower individuals to make informed decisions regarding their data and interactions with tools that track their activity.
\section{Limitations and Future Work}
We acknowledge that our study has several limitations. First, the ongoing debate surrounding the classification of gig workers in the U.S. \cite{Cherry_2023, Scheiber_2022} may affect the relevance of our findings if \textcolor{black}{their classification would change in any way}. A shift in classification could alter gig workers’ accountabilities and their use of self-tracking tools, as we detailed in section 5.2. Second, our study focused on the experiences of gig drivers in the U.S., so the findings may not completely apply to other gig workers (i.e., freelance workers, Airbnb hosts), and they may also vary depending on the legal responsibilities of workers in different countries. For example, variations in tax regulations may affect the relevance of the section on tax-related accountabilities. Future work can explore self-tracking in other gig work contexts, especially those significantly different from gig drivers or in other countries. Third, although we made efforts to diversify our participant pool by recruiting both online and offline, a portion of our participants found our study through Reddit, potentially favoring users with a heightened awareness of gig worker-related accountabilities and self-tracking practices. Lastly, as our study is qualitative, we encourage further investigation into gig workers' self-tracking through alternative research methods and with larger sample sizes.
\section{Conclusion}
Our study has revealed how gig workers self-track to effectively manage the multiple accountabilities associated with their various identities. We connect this practice to neoliberalism, highlighting the unpaid self-tracking labor that gig workers undertake to sustain their employment. While acknowledging self-tracking’s supportive role, it's crucial to emphasize its burdensome necessity for gig workers to thrive and continue working in the gig economy. Furthermore, we explore how gig workers self-track to address inherent imbalances in the gig economy which often hinder their success. This is evident in their struggles with unfair compensation and the extensive effort required to regain autonomy over their professional lives, which are managed by self-tracking. Finally, we propose design considerations for platforms to better support workers’ multidimensional tracking needs.
\begin{acks}
    We extend our sincere appreciation to the anonymous reviewers whose constructive feedback significantly contributed to the refinement and improvement of our manuscript. Our gratitude extends to the 25 interview participants for their support and participation in this study. Special thanks are also extended to the owners, managers, and employees of local establishments, and subreddit moderators who graciously allowed us to promote our research. Their cooperation has been instrumental in the success of our study.
\end{acks}

\bibliographystyle{ACM-Reference-Format}
\bibliography{references}

\end{document}